\documentclass[journal,final,twocolumns,10pt]{IEEEtran}
\IEEEoverridecommandlockouts

\usepackage{cite}
\usepackage{indentfirst}
\usepackage{graphicx}
\usepackage{changepage}
\usepackage{stfloats}
\usepackage{amsfonts,amssymb}
\usepackage{bm}
\usepackage{algorithm}
\usepackage{algorithmic}
\usepackage{amsmath}
\usepackage{amsmath,amssymb,amsfonts}
\usepackage{times}
\usepackage{url}
\usepackage{textcomp}
\usepackage{xcolor}
\usepackage{color}
\usepackage{setspace}
\usepackage{enumitem}
\usepackage{float}
\usepackage{diagbox}
\usepackage[T1]{fontenc}
\usepackage[utf8]{inputenc}
\usepackage{authblk}
\usepackage{threeparttable}
\usepackage{booktabs}
\usepackage{footnote}
\usepackage{graphicx}
\usepackage{pifont}
\usepackage{subfigure}
\usepackage{mathrsfs}

\renewcommand{\theenumi}{\alph{enumi}}

\DeclareMathOperator*{\argmin}{arg\,min}

\newtheorem{remark}{Remark}

\begin{document}


\title{Sensing for Reliable UAV Communication: Robust Trajectory and Resource Optimization in Low-Altitude Networks}

\author{\IEEEauthorblockN{Yifan~Jiang, Qingqing~Wu, \textit{Senior Member, IEEE}, Hongxun~Hui, \textit{Member, IEEE}, Wen~Chen, \textit{Senior Member, IEEE}, Wei~Feng, \textit{Senior Member, IEEE}, Shanpu~Shen, \textit{Senior Member, IEEE}}

\thanks{Yifan~Jiang is with Shanghai Jiao Tong University, Shanghai 200240, China, and also with the State Key Laboratory of Internet of Things for Smart City, University of Macau, Macao 999078, China (email: yc27495@umac.mo). 
Qingqing~Wu and Wen~Chen are with the Department of Electronic Engineering, Shanghai Jiao Tong University, Shanghai 200240, China (e-mail: \{qingqingwu@sjtu.edu.cn; wenchen@sjtu.edu.cn\}).
Hongxun~Hui and Shanpu~Shen are with the State Key Laboratory of Internet of Things for Smart City and Department of Electrical and Computer Engineering, University of Macau, Macao, 999078 China (email: \{hongxunhui@um.edu.mo; shanpushen@um.edu.mo\}).
Wei~Feng is with Department of Electronic Engineering, State Key Laboratory of Space Network and Communications, Tsinghua University, Beijing 100084, China (e-mail: fengwei@tsinghua.edu.cn).}}

\maketitle

\begin{abstract} 
    In low-altitude wireless networks, sensing-aided communication has emerged as a promising integrated sensing and communication (ISAC) paradigm for unmanned aerial vehicle (UAV) tracking and communication. 
    This paper investigates reliable sensing-aided communication for multiple cellular-connected UAVs under mobility uncertainties. 
    Specifically, we maximize the minimum outage capacity among UAVs by jointly optimizing their real-time predicted positions, as well as the base station (BS) transmit power and bandwidth allocations.
    To address the non-convex and intractable maximum tolerable outage probability (OP) constraints, two robust optimization schemes are proposed based on a continuous confidence ellipse (CE) and discretized inverse-whitened sectors (IWSs), respectively.     
    For the CE-based scheme, an efficient algorithm is proposed to optimize the predicted UAV positions individually via block successive convex approximation, followed by convex resource allocation. 
    For the IWS-based scheme, an IWS-based OP approximation is proposed to facilitate the robust optimization, based on which a low-complexity IWS selection method is proposed to decouple the optimization variables. 
    Then, a similar sequential optimization algorithm is proposed based on the projected gradient descent approach.
    The two algorithms are further unified into a common trajectory-resource optimization framework, revealing a low-complexity structure for robust UAV trajectory and resource management.
    Simulation results validate the effectiveness of our proposed OP approximation, demonstrate the significant outage capacity improvement of the proposed robust optimization schemes over benchmark schemes, and illustrate the superiority of the IWS-based scheme over the CE-based scheme. 
\end{abstract}


\section{Introduction}\label{Sec:I}

In recent years, the unprecedented proliferation of unmanned aerial vehicle (UAV)-enabled applications in logistics, surveillance, agriculture, and other emerging scenarios has revealed significant economic potential and promoted the development of the low-altitude economy \cite{LAE}. 
To accommodate the widespread and diverse UAV applications, the low-altitude wireless network (LAWN) has emerged as a promising network architecture integrating sensing, communication, and other foundational functionalities \cite{LAWN,JinCO}. 
Therefore, integrated sensing and communication (ISAC), also as one of the key enabling technologies for the upcoming sixth-generation networks, has been anticipated to support both sensing and communication services for users in LAWNs \cite{LAWN,zzy-LA,IRS-ISAC,ISAC2,qiaoyan-Sens}. 
To be specific, high-performance sensing and communication can be achieved through the efficient utilization of shared spectrum and hardware resources in ISAC. 
Moreover, ISAC enables the mutual assistance between sensing and communication, thereby considerably enhancing the system performance over conventional dedicated sensing or communication systems. 

To enhance the system performance in typical ISAC scenarios such as simultaneous user tracking and communication, sensing-aided/assisted communication has emerged as a promising paradigm, as it not only enables the dual functions but also improves the communication performance by efficiently leveraging sensing information \cite{LAWN,IRS-ISAC,PB1,Relia2,LAE4,S-Handover,RA-HO,covert}.
Specifically, the echoes of ISAC signals transmitted toward the users are collected by ISAC base stations (BSs) to obtain sensing information, which is then utilized to assist communication design. 
For example, in high-mobility scenarios where communication users are also tracked by ISAC BSs, the user directions predicted from sensing information can be utilized for predictive beam tracking \cite{PB1}.
In this manner, the feedback overhead incurred in conventional communication schemes can be substantially reduced, which is particularly favorable for real-time tracking. 
Also, sensing information can be exploited to achieve accurate beam alignment and adaptive beamwidth control \cite{Relia2}, thereby significantly improving user data rates. 
Besides, when users move across cells in wireless cellular networks, sensing information can be leveraged to design handover strategies for flexible wireless coverage \cite{LAE4,S-Handover,RA-HO}. 
Additionally, the sensing-aided communication paradigm can also be applied in secure communication scenarios, where sensing information obtained by tracking adversary targets is utilized to improve the communication security for legitimate users \cite{covert}.

Apart from the aforementioned techniques, sensing information can also be exploited to construct sensing performance metrics incorporated in resource allocation among multiple users under the sensing-aided communication paradigm \cite{PB1,fwd,PCRB-MUI,secure-MEC,Game,OFDM-RA,coo-RA}.
For instance, the posterior Cramér-Rao bound (PCRB) was derived from predicted user positions and velocities in \cite{PB1} to characterize the accuracy for multi-user tracking and then served as the minimization objective for power allocation, thereby improving both tracking accuracy and communication data rates over conventional schemes.
In \cite{fwd}, the PCRB was treated as the tracking quality-of-service (QoS) for ISAC users and minimized by optimizing the bandwidth and power allocation under a unified framework for user detection, localization, and tracking while satisfying communication QoS requirements. 
To address the inter-user sensing interference, a power allocation scheme was proposed in \cite{PCRB-MUI} to maximize the sum-rate of users subject to maximum PCRB constraints for each user, which effectively improves the overall communication performance. 
In addition, to facilitate secure communication, the joint optimization of power allocation and sensing signal covariance matrix for were studied in \cite{secure-MEC}, subject to constraints on the maximum eavesdropping rate and sensing error of the mobile eavesdropper. 
In \cite{Game}, an optimization framework for transmit beamforming and power allocation was proposed to minimize the total network power consumption subject to constraints on a minimum secrecy rate among users and a maximum sensing error for the adversary target.
Besides the monostatic ISAC scenario in \cite{PB1,fwd,PCRB-MUI,secure-MEC,Game}, resource allocation for sensing-aided communication in the multi-static scenario was studied in \cite{OFDM-RA,coo-RA}. 
In \cite{OFDM-RA}, the orthogonal frequency division multiplexing signals transmitted to users were received by multiple BSs and the subcarrier and power allocation were optimized for maximizing the weighted sum-rate of users subject to maximum PCRB constraints. 
To further enhance sensing performance via multiple BS cooperation, the BS assignment, bandwidth, and power allocation were jointly optimized to maximize a overall performance metric incorporating both sum-PCRB and sum-rate of users \cite{coo-RA}.

The aforementioned studies focused on tracking standalone users or targets, in which case the user or target movement cannot participate in the overall system design.
By contrast, the trajectories of cellular-connected UAVs can be designed through real-time remote control from BSs to maintain reliable BS-UAV communication links, typically for applications such as emergency rescue and remote inspection \cite{qqw2021JSAC,YZ2019PIEEE}. 
Nevertheless, due to random environmental variations and hardware imperfections in practice, such as wind gusts, jittering, and control errors \cite{2020TCOM-Uncertain,ISAC-UAV-2}, the ground-truth UAV movement is in fact unknown to BSs and the real-time UAV trajectories can only be partially designed, which severly degrades the system stability. 
To handle this issue, the sensing-aided communication paradigm can be employed, where ISAC signals are transmitted from BSs to cellular-connected UAVs not only for communication but also to acquire useful statistical information about UAV's movement via tracking. 
For example, the popular extended Kalman filtering (EKF) technique can be applied to accurately track UAV's movement and obtain the mean square error (MSE) for the tracking result \cite{PB1,Relia1}.
In this case, the beam alignment between BSs and UAVs need to be sufficiently robust to combat with the tracking error for reliable communication, while the beam alignment accuracy is highly dependent on UAV's position/trajectory \cite{Relia2,covert,robustBF}. 
Since the tracking errors can be time-varying during UAV's flight, it is crucial to study robust real-time UAV trajectory optimization for reliable sensing-aided communication. 
In multi-UAV scenarios, the reliable communication performance of the overall system depends on both trajectory design and resource allocation of cellular-connected UAVs, posing challenges to address such coupling. 
Therefore, it is worth investigating the robust trajectory and resource optimization for multiple cellular-connected UAVs under the sensing-aided communication paradigm, which also remains uninvestigated and thus motivates our work.

In this paper, we investigate the robust optimization of multiple cellular-connected UAV trajectories and their resource allocation for reliable sensing-aided communication in low-altitude wireless networks. 
Specifically, the trajectories of cellular-connected UAVs are proactively designed in real time through remote control from a BS while influenced by uncertain environmental variations or hardware imperfections. 
For the signal reception reliability of the UAVs, the BS simultaneously tracks and communicates with the multiple UAVs in orthogonal frequency bands under the sensing-aided communication paradigm. 
To ensure real-time UAV tracking, the predictive beamforming is adopted to reduce the system signaling overhead and the resulting reliable spectral efficiency of each UAV is characterized by the outage capacity. 
The main contributions of this paper are summarized as follows:
\begin{itemize}
    \item A reliable sensing-aided communication scheme is proposed for multi-UAV simultaneous tracking and communication, where the minimum outage capacity among multiple UAVs is maximized for the fairness of UAVs by optimizing the predicted UAV positions and the allocations of BS transmit power and bandwidth. 
    The communication reliability of each UAV is guaranteed by a maximum tolerable outage probability (OP). 
    \item To handle the intractable maximum tolerable OP constraints, two robust optimization schemes are proposed by constructing an uncertainty set of the random ground-truth UAV positions based on a confidence ellipse (CE) and discretized sectors obtained via the inverse whitening transform, respectively. 
    The solution robustness is guaranteed by constraining that a complementary outage region (COR) characterizing the OP contains the constructed CE and inverse-whitened sector (IWSs). 
    Particularly, an effective IWS-based OP approximation is proposed to facilitate the robust optimization. 
    \item For the CE-based scheme, we develop an efficient sequential algorithm that first optimizes each UAV’s predicted position via block successive convex approximation (BSCA), and then solves a convex resource allocation subproblem. 
    Inspired by this structure, we further design a low-complexity IWS selection strategy to decouple trajectory optimization from resource allocation in the IWS-based scheme, where each predicted UAV position is optimized via projected gradient descent (PGD). 
    The two schemes are finally unified into a common sequential optimization framework, offering useful insights into low-complexity robust trajectory and resource optimization through proper uncertainty set construction.
    \item Numerical results validate the effectiveness of the proposed IWS-based OP approximation and the proposed robust optimization schemes. 
    Compared to the benchmark schemes dedicated for sensing or communication, both of our proposed robust optimization schemes achieve significant outage capacity improvement. 
    Besides, the IWS-based scheme is more effective than the CE-based scheme for robust trajectory and resource optimization. 
\end{itemize}

\emph{Notation:} $\mathbf{1}_{m}$ denote a $m\times 1$ column vector with all elements equal to 1, respectively. 
$\mathbf{I}_{m}$ denotes the $m\times m$ identity matrix. 
$\mathcal{O}(\cdot)$ represents the big-O notation for computational complexity. 
$\mathbb{E}[\cdot]$ denotes the statistical expectation. 
$\bm{\mathcal{N}}(\mathbf{0},\mathbf{Q}^{\text{p}})$ denotes a real-valued Gaussian distribution with a mean vector $\mathbf{x}$ and covariance matrix $\mathbf{Q}$ and $\sim$ means ``distributed as''. 
$\otimes$ is the Kronecker product. 
$\mathrm{diag}(b_{1},...,b_{L})$ denotes a diagonal matrix with $b_{1},...,b_{L}$ being its diagonal elements. 
For an arbitrary matrix $\mathbf{A}$, $\mathbf{A}^{T}$, $\mathbf{A}^{-1}$, $\mathrm{det}(\mathbf{A})$, and $[\mathbf{A}]_{ij}$ denote its transpose, inverse, determinant, and $(i,j)$-th element, respectively. 
$\nabla_{x} f(x;y)$ represents the gradient of the function $f(x;y)$ with respect to the variable $x$.

\vspace{-1mm}
\section{System Model}\label{Sec:II}

\begin{figure}[!t]
    \centering
    \includegraphics[width=0.38\textwidth]{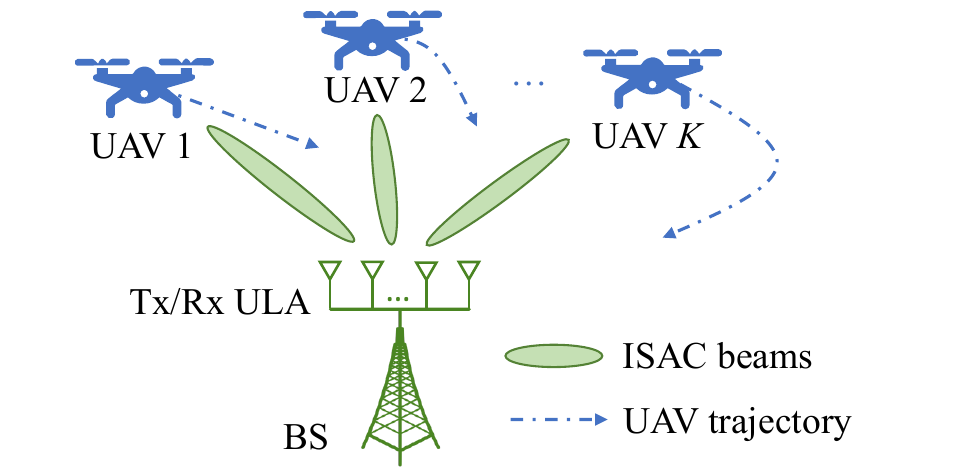}
    \caption{A low-altitude ISAC system.}
    \label{fig0}
    \vspace{-3mm}
\end{figure}

As shown in Fig. \ref{fig0}, we consider a downlink low-altitude ISAC system where a terrestrial BS simultaneously tracks and communicates with $K$ cellular-connected UAVs using ISAC signals over a service duration of $T$ s. 
Specifically, the movements of cellular-connected UAVs are remotely controlled by the BS in real time \cite{qqw2021JSAC,qqw2024PIEEE,YFJ2026TWC}. 
However, the ground-truth UAV motion states (i.e., positions and velocities) are random and cannot be perfectly known by the BS due to practical environmental factors and hardware imperfections \cite{2020TCOM-Uncertain}. 
Therefore, the BS leverages sensing results from echo signals for UAV tracking and precise beam alignment.
To guarantee real-time UAV tracking and avoid inter-user interference, the BS assigns orthogonal frequency resources to each UAV \cite{qqw2021JSAC}. 
As an initial study, we assume that the BS is equipped with uniform linear arrays (ULAs) consisting of $N_{\rm{t}}$ transmit antennas and $N_{\rm{r}}$ receive antennas, and each UAV is equipped with a single antenna.\footnote{The antenna setting can be extended to more sophisticated cases. For example, in the case with planar arrays, the antenna array beamforming gain can be modeled as a product of antenna array gains resulting from two ULAs.} 
Moreover, the $k$-th UAV is assumed to fly at a fixed altitude of $H_{k}$ meters\footnote{The UAV movement can be readily extended to the three-dimensional case by incorporating the vertical movement into the motion state.} with $k\in\mathcal{K} = \{1,2,\ldots,K\}$, and thus a two-dimensional Cartesian coordinate system is adopted to represent the horizontal positions of the BS and UAVs. 
Without loss of generality, the BS is assumed to be located at the origin. 

\subsection{UAV Mobility Model}

In this paper, we adopt the standard assumption that the motion states (i.e., positions and velocities) of UAVs remain constant within a sufficiently short time interval $\Delta T$ \cite{qqwMultiUAV}. 
Accordingly, the service duration $T$ is discretized into $N = T/\Delta T$ time slots indexed by $n \in \mathcal{N}_{\rm{T}} = \{1,2,\ldots,N\}$.
For the $k$-th UAV, its motion state at the $n$-th time slot is defined as
$\mathbf{x}_{k,n} \triangleq [x_{k,n}, v_{k,n}^{\text{x}}, y_{k,n}, v_{k,n}^{\text{y}}]^T, k \in \mathcal{K}$, where $x_{k,n}$, $v_{k,n}^{\text{x}}$, $y_{k,n}$, and $v_{k,n}^{\text{y}}$ denote the $x$-axis coordinate, velocity along the $x$-axis, $y$-axis coordinate, and velocity along the $y$-axis, respectively.
Based on the standard discrete-time control model \cite{YFJ2025TVT}, the mobility model of the $k$-th UAV is expressed as
\vspace{-1.5mm}
\begin{equation}
    \mathbf{x}_{k,n} = \mathbf{G}\mathbf{x}_{k,n-1} + \mathbf{u}_{k,n} + \mathbf{z}_{k,n}^{\text{p}}, \quad \forall k\in\mathcal{K}, n\in\mathcal{N}_{\rm{T}}, \label{fm:Mobi} 
    \vspace{-1.5mm}
\end{equation}
where $\mathbf{G}\in\mathbb{R}^{4\times 4}$ is the transition matrix, $\mathbf{u}_{k,n}\in\mathbb{R}^{4\times 1}$ represents the control input from the BS, and $\mathbf{z}_{k,n}^{\text{p}} \sim \bm{\mathcal{N}}(\mathbf{0},\mathbf{Q}^{\text{p}})$ denotes the process noise that models random disturbances \cite{YFJ2025TVT}. 
The expressions of the transition matrix $\mathbf{G}$ and the process noise covariance matrix $\mathbf{Q}^{\text{p}} = \mathbb{E}[\mathbf{z}_{k,n}^{\text{p}}(\mathbf{z}_{k,n}^{\text{p}})^{T}]$ can be obtained from the classic constant-velocity model as \cite{YFJ2024CL}
\vspace{-1.5mm}
\begin{equation}
    \mathbf{G} = \mathbf{I}_{2} \otimes \begin{bmatrix}
        1 & \Delta T  \\
        0 & 1 
    \end{bmatrix}, 
    \mathbf{Q}^{\text{p}} = \mathbf{I}_{2} \otimes \begin{bmatrix}
        \frac{1}{3}\Delta T^{3} & \frac{1}{2}\Delta T^{2}  \\
        \frac{1}{2}\Delta T^{2} & \Delta T 
    \end{bmatrix}\tilde{q},
    \vspace{-1.5mm}
\end{equation}
where $\tilde{q}$ denotes the process noise intensity \cite{YFJ2024CL}.
In particular, the UAV mobility model (\ref{fm:Mobi}) can be interpreted as introducing the external motion control input from the BS to refine the UAV motion states from a default constant-velocity motion. 

Although the ground-truth UAV motion states are unavailable to the BS, the BS can obtain their estimates through sensing at each time slot, denoted by 
\vspace{-1.5mm}
\begin{equation}
    \hat{\mathbf{x}}_{k,n} = [\hat{x}_{k,n}, \hat{v}_{k,n}^{\text{x}}, \hat{y}_{k,n}, \hat{v}_{k,n}^{\text{y}}]^{T}, \quad \forall k\in\mathcal{K}, n\in\mathcal{N}_{\rm{T}}. 
\vspace{-1.5mm}
\end{equation} 
As such, at each time slot, the BS can design the motion control input and predict the $k$-th UAV's motion state at the next time slot based on the estimated UAV motion states. 
Specifically, the predicted motion state of the $k$-th UAV at the $n$-th time slot from the $(n-1)$-th time slot is denoted by 
\vspace{-1.5mm}
\begin{equation}
    \!\!\mathbf{x}_{k,n|n-1} \!=\! [x_{k,n|n-1},\!v_{k,n|n-1}^{\text{x}},\!y_{k,n|n-1},\!v_{k,n|n-1}^{\text{y}}]^{T}.
\vspace{-1.5mm}
\end{equation} 
Then, the BS controls the $k$-th UAV's flight by designing the motion control input as $\mathbf{u}_{k,n} = \mathbf{x}_{k,n|n-1} - \mathbf{G}\hat{\mathbf{x}}_{k,n-1}$. 
Consequently, the mobility model (\ref{fm:Mobi}) can be reformulated as 
\vspace{-1.5mm}
\begin{equation}
    \mathbf{x}_{k,n} = \mathbf{x}_{k,n|n-1} + \mathbf{G}(\mathbf{x}_{k,n-1} - \hat{\mathbf{x}}_{k,n-1}) + \mathbf{z}_{k,n}^{\text{p}}, \label{fm:pre}
    \vspace{-1.5mm}
\end{equation}
which indicates that the $k$-th UAV's trajectory during the service duration, denoted by $\{\mathbf{q}_{k,n}\}, n\in\mathcal{N}_{\rm{T}}$, can be partially designed online by optimizing the predicted UAV position $\mathbf{q}_{k,n|n-1} = [x_{k,n|n-1}, y_{k,n|n-1}]^{T}$ at each time slot, where $\mathbf{q}_{k,n} = [x_{k,n}, y_{k,n}]^{T}$ denotes the ground-truth UAV position. 
In addition, it can be noted from (\ref{fm:pre}) that the prediction error arises from the estimation error propagated from the $(n-1)$-th time slot and the process noise. 

\vspace{-1mm}
\subsection{Sensing Model}

At each time slot, the BS transmits ISAC signals and measures the azimuth angles and distances of the UAVs from the echo signals. 
To ensure uninterrupted beam coverage for UAVs, the BS employs the low-overhead predictive beamforming for ISAC signal transmission based on the predicted UAV motion states \cite{YFJ2026TWC}. 
As a result, the BS transmit beamforming vector for the $k$-th UAV at the $n$-th time slot is given by 
\vspace{-1.5mm}
\begin{equation}
    \mathbf{f}_{k,n|n-1} = \mathbf{a}(\theta_{k,n|n-1}) = \mathbf{a}\left(\arctan\left(\frac{y_{k,n|n-1}}{x_{k,n|n-1}}\right)\right), 
    \vspace{-1.5mm}
\end{equation}
where $\mathbf{a}(\cdot)$ denotes the transmit steering vector of the BS and $\theta_{k,n|n-1}$ denotes the predicted azimuth angle. 
The expression of $\mathbf{a}(\cdot)$ is given by 
\vspace{-1.5mm}
\begin{equation}
    \mathbf{a}(\theta) = \frac{1}{\sqrt{N_{\rm{t}}}} \left[ 1, ..., \mathrm{exp}(-j\pi(N_{\rm{t}}-1)\cos{\theta})\right]^{T}.
    \vspace{-1.5mm}
\end{equation}

In this paper, the ISAC signals are assumed to propagate through line-of-sight (LoS)-dominant channels and experience free-space path loss due to the limited signal blockages and scatterings in the vertical dimension \cite{qqwMultiUAV}. 
After receiving the echo signals from the $k$-th UAV over its allocated frequency band, the BS applies the matched-filtering technique to obtain the measured azimuth angle and distance \cite{KTMeng2024TVT-IS}, denoted by $\hat{\theta}_{k,n}$ and $\hat{d}_{k,n}$, respectively. 
Thus, by defining the measured result vector $\bm{\omega}_{k,n} \triangleq [\hat{\theta}_{k,n}, \hat{d}_{k,n}]^{T}$, the relationship between the measured and ground-truth results can be represented by a measurement model formulated as 
\vspace{-1.5mm}
\begin{equation}
    \bm{\omega}_{k,n} 
    = \mathbf{h}(\mathbf{x}_{k,n}) + \mathbf{z}_{k,n}^{\text{m}}   
    = \begin{bmatrix}
        \theta_{k,n} \\
        d_{k,n}
    \end{bmatrix}
    + \begin{bmatrix}
        z_{k,n}^{\text{az}} \\
        z_{k,n}^{\text{dst}}
    \end{bmatrix}\!, \label{fm:mea}
    \vspace{-1.5mm}
\end{equation}
where $\theta_{k,n}$ and $d_{k,n}$ denote the ground-truth azimuth angle and distance of the $k$-th UAV given by  
\vspace{-1.5mm}
\begin{equation}
    \theta_{k,n} 
    = \arctan(y_{k,n}/x_{k,n}), \
    d_{k,n} 
    = \sqrt{\|\mathbf{q}_{k,n}\|^{2} + H_{k}^{2}},
    \vspace{-1.5mm}
\end{equation}
respectively, and $\mathbf{z}_{k,n}^{\text{m}} \sim \bm{\mathcal{N}}(\bm{0},\mathbf{Q}_{k,n}^{\rm{m}})$ represents the measurement noise vector. 
The covariance matrix for the measurement noise is given by $\mathbf{Q}_{k,n}^{\rm{m}} = \mathrm{diag}((\sigma_{k,n}^{\text{az}})^{2}, (\sigma_{k,n}^{\text{dst}})^{2})$ with \cite{YFJ2024CL}
\vspace{-1.5mm}
\begin{equation}
    (\sigma_{k,n}^{\text{az}})^{2} = \frac{ a^{\text{az}} \sigma_{\rm{r}}^{2} d_{k,n}^{2}  }{ p_{k,n} \rho_{\text{s}} \sin(\theta_{k,n})^{2}},  \quad 
    (\sigma_{k,n}^{\text{dst}})^{2} = \frac{ a^{\text{dst}} \sigma_{\rm{r}}^{2} d_{k,n}^{2} }{ p_{k,n} \rho_{\text{s}} b_{k,n} }. \label{fm:sgm}
    \vspace{-1.5mm}
\end{equation}
In (\ref{fm:sgm}), $a^{\text{az}}$ and $a^{\text{dst}}$ represent the coefficients for azimuth-angle and distance measurements, respectively, which are dependent on specific system configurations and signal processing techniques \cite{YFJ2024CL}. 
$\sigma_{\rm{r}}^{2}$ denotes the additive white Gaussian noise power at the BS receiver, while $p_{k,n}$ and $b_{k,n}$ denote the transmit power and bandwidth allocated by the BS to the $k$-th UAV, respectively. 
The coefficient $\rho_{\text{s}}$ is given by 
\vspace{-1.5mm}
\begin{equation}
    \rho_{\text{s}} = \frac{\lambda^{2}\sigma_{\rm{rcs}}G_{\rm{ar}}G_{\rm{mf}}}{(4\pi)^{3}}, 
    \vspace{-1.5mm}
\end{equation}
where $\lambda$, $\sigma_{\rm{rcs}}$, $G_{\rm{ar}}$, and $G_{\rm{mf}}$ denote the signal carrier wavelength, UAV radar cross-section, BS antenna array gain, and matched-filtering gain \cite{KTMeng2024TVT-IS}, respectively.

Due to the non-linear measurement model (\ref{fm:mea}), the BS estimates the motion states of UAVs following the standard EKF procedures \cite{MKay}. 
Specifically, the estimated motion state of the $k$-th UAV at the $n$-th time slot is given by 
\vspace{-1.5mm}
\begin{equation}
    \hat{\mathbf{x}}_{k,n} = \mathbf{x}_{k,n|n-1} + \mathbf{K}_{k,n}(\bm{\omega}_{k,n} - \mathbf{h}(\mathbf{x}_{k,n|n-1})), \label{fm:hat}
\vspace{-1.5mm}
\end{equation} 
where $\mathbf{K}_{k,n}$ denotes the Kalman gain matrix given by 
\vspace{-1.5mm}
\begin{equation}
    \mathbf{K}_{k,n} = \mathbf{M}_{k,n}^{\text{p}}\mathbf{H}_{k,n}^{T}(\mathbf{Q}_{k,n}^{\text{m}} + \mathbf{H}_{k,n}\mathbf{M}_{k,n}^{\text{p}}\mathbf{H}_{k,n}^{T} )^{-1}. \label{fm:Kn}
\vspace{-1.5mm}
\end{equation}  
In (\ref{fm:Kn}), $\mathbf{M}_{k,n}^{\text{p}}$ denotes the prediction MSE matrix and $\mathbf{H}_{k,n}\in \mathbb{R}^{2\times 4}$ denotes the Jacobian matrix of the function $\mathbf{h}(\cdot)$ in the measurement model (\ref{fm:mea}).
The matrices $\mathbf{M}_{k,n}^{\text{p}}$ and $\mathbf{H}_{k,n}$ are given by 
\vspace{-1.5mm}
\begin{align}
    \mathbf{M}_{k,n}^{\text{p}} &= \mathbf{G}\mathbf{M}_{k,n-1}\mathbf{G}^{T} + \mathbf{Q}^{\text{p}}, \label{fm:Mpn} \\
    \mathbf{H}_{k,n} &= \frac{\partial\mathbf{h}}{\partial\mathbf{x}_{k,n}}\Big|_{\mathbf{x}_{k,n} = \mathbf{x}_{k,n|n-1}}, \label{fm:Hn}
\vspace{-1.5mm}
\end{align}
respectively, where $\mathbf{M}_{k,n-1}$ denotes the state estimation MSE matrix at the $(n-1)$-th time slot. 
According to \cite{YFJ2026TWC}, $\mathbf{x}_{k,n} \sim \bm{\mathcal{N}}(\mathbf{x}_{k,n|n-1},\mathbf{M}_{k,n}^{\rm{p}})$ and $\mathbf{x}_{k,n} \sim \bm{\mathcal{N}}(\hat{\mathbf{x}}_{k,n},\mathbf{M}_{k,n})$ both approximately hold, and the state estimation MSE matrix at the $n$-th time slot $\mathbf{M}_{k,n}$ is given by $\mathbf{M}_{k,n} = (\mathbf{I}_{4} - \mathbf{K}_{k,n}\mathbf{H}_{k,n})\mathbf{M}_{k,n}^{\text{p}}$.

\subsection{Communication Model}

Given the LoS-dominant channel with free-space path loss, the received signal-to-noise ratio (SNR) of the $k$-th UAV at the $n$-th time slot is expressed as\footnote{In the sensing-aided predictive beamforming scheme, the BS transmit beamformer can be designed based on either the predicted or estimated motion state \cite{YFJ2026TWC}. However, since the estimated motion state is usually more precise, this paper focuses on the communication performance resulting from the prediction-based beamforming as a conservative evaluation of reliable communication rate.} 
\vspace{-1.5mm}
\begin{equation}
    \gamma_{k,n} = \frac{p_{k,n}\beta_{\rm{c}}N_{\rm{t}}\delta(\mathbf{q}_{k,n|n-1},\mathbf{q}_{k,n})}{ b_{k,n} N_{\rm{c}} (\|\mathbf{q}_{k,n}\|^{2} + H_{k}^{2}) }, \label{fm:snr} 
    \vspace{-1.5mm}
\end{equation}
where $\beta_{\rm{c}} = (\lambda/4\pi)^{2}$ denotes the channel power gain at the reference distance of 1 m, $N_{\rm{c}}$ denotes the thermal noise power spectral density at the UAV receiver, and $\delta(\cdot)$ denotes the predictive beam alignment factor \cite{qqwMultiUAV,YFJ2026TWC}. 
The specific expression of $\delta(\mathbf{q}_{k,n|n-1},\mathbf{q}_{k,n})$ is given by 
\vspace{-1.5mm}
\begin{equation}
    \delta(\mathbf{q}_{k,n|n-1},\mathbf{q}_{k,n}) = \Bigl|\frac{\sin(\frac{N_{\rm{t}}\pi}{2}\kappa(\mathbf{q}_{k,n|n-1},\mathbf{q}_{k,n}))}{N_{\rm{t}}\sin(\frac{\pi}{2}\kappa(\mathbf{q}_{k,n|n-1},\mathbf{q}_{k,n}))}\Bigr|^{2}, 
    \vspace{-1.5mm}
\end{equation}
where the function $\kappa(\cdot)$ is defined by 
\vspace{-1.5mm}
\begin{equation}
    \kappa(\mathbf{q}_{k,n|n-1},\mathbf{q}_{k,n})
    \triangleq \frac{x_{k,n|n-1}}{\|\mathbf{q}_{k,n|n-1}\|} - \frac{x_{k,n}}{\|\mathbf{q}_{k,n}\|}.
    \vspace{-1.5mm}
\end{equation} 
Given (\ref{fm:snr}), the achievable rate of the $k$-th UAV at the $n$-th time slot is given by $R_{k,n} = b_{k,n}\log_{2}( 1 + \gamma_{k,n} )$.

Note that the ISAC signals received by the $k$-th UAV suffer from slow fading owing to the random but invariant UAV position at each time slot \cite{YFJ2026TWC}. 
Therefore, the reliable communication rate of the $k$-th UAV is characterized by its outage capacity denoted by $\bar{R}_{k,n}$, which refers to the maximum achievable rate such that the $k$-th UAV's outage probability (OP) is less than a maximum tolerable OP threshold denoted by $\epsilon_{k}$ \cite{goldsmith}. 
The $k$-th UAV's OP at the $n$th time slot is represented by 
\vspace{-1.5mm}
\begin{equation}
    \zeta_{k,n}(\bar{R}) = \mathbb{P}( R_{k,n} < \bar{R}), 
    \vspace{-1.5mm}
\end{equation}
where $\bar{R}$ denotes a target rate.

\subsection{Problem Formulation}

Let $\mathbf{p}_{n} = [p_{1,n},\ldots,p_{K,n}]^{T}$ and $\mathbf{b}_{n} = [b_{1,n},\ldots,b_{K,n}]^{T}$ denote the vector for BS transmit power and bandwidth allocation at the $n$-th time slot, respectively. 
To ensure the fairness and communication reliability among UAVs, a reliable communication scheme is proposed to maximize the minimum outage capacity among all UAVs by optimizing the predicted UAV positions $\{\mathbf{q}_{k,n|n-1}\}, k\in\mathcal{K}$, BS transmit power allocation $\mathbf{p}_{n}$, and bandwidth allocation $\mathbf{b}_{n}$ at the $n$-th time slot. 
The optimization problem of the proposed reliable communication scheme is formulated as
\vspace{-1.5mm}
\begin{align}
    (\mathrm{P1}): \ &\max_{ \mathbf{p}_{n}, \mathbf{b}_{n}, \{\mathbf{q}_{k,n|n-1}\} } \ \min_{k\in\mathcal{K}} \ \bar{R}_{k,n}  \label{opt-obj} \\
    \text{s.t.} \ 
    &p_{\rm{min}} \leq p_{k,n} \leq p_{\rm{max}}, \forall k, \tag{\ref{opt-obj}{a}} \label{opt-cstrt-a} \\
    &b_{\rm{min}} \leq b_{k,n} \leq b_{\rm{max}}, \forall k, \tag{\ref{opt-obj}{b}} \label{opt-cstrt-b} \\
    &\mathbf{1}_{K}^{T}\mathbf{p}_{n} = p_{\rm{tot}}, \mathbf{1}_{K}^{T}\mathbf{b}_{n} = b_{\rm{tot}}, \tag{\ref{opt-obj}{c}} \label{opt-cstrt-c} \\
    &\|\mathbf{q}_{k,n|n-1} - \hat{\mathbf{q}}_{k,n-1} \| \leq V_{\rm{max}}\Delta T, \forall k, \tag{\ref{opt-obj}{d}} \label{opt-cstrt-d} \\
    &\zeta_{k,n}(\bar{R}_{k,n}) < \epsilon_{k}, \forall k, \tag{\ref{opt-obj}{e}} \label{opt-cstrt-e}
    \vspace{-1.5mm}
\end{align}
where $\hat{\mathbf{q}}_{k,n-1} = [\hat{x}_{k,n-1}, \hat{y}_{k,n-1}]^{T}$ denotes the estimated position of the $k$-th UAV at the $(n-1)$-th time slot, and $V_{\rm{max}}$ denotes the maximum UAV velocity. 
In (P1), (\ref{opt-cstrt-a})-(\ref{opt-cstrt-c}) denote the BS transmit power and bandwidth budget constraints, (\ref{opt-cstrt-d}) denotes the UAV mobility constraint, and (\ref{opt-cstrt-e}) represents the maximum tolerable OP constraint. 
(P1) is non-trivial to be optimally solved due to the coupling among optimization variables in the non-convex probabilistic constraint (\ref{opt-cstrt-e}). 

\section{Proposed Robust Optimization Schemes}

To address the intractable probabilistic constraint (\ref{opt-cstrt-e}) in (P1), the OP $\zeta_{k,n}(\bar{R}_{k,n})$ is first approximated via Taylor expansions and the integral of the ground-truth UAV position probability density function (PDF) over a set termed the COR. 
Then, two robust optimization schemes are proposed by constructing an uncertainty set of possible ground-truth UAV positions based on either a continuous CE or discretized sectors obtained after the inverse whitening transform \cite{whitening0}, referred to as the CE-based and IWS-based schemes, respectively.
The constructed uncertainty set is constrained in the COR to satisfy the maximum tolerable OP constraint (\ref{opt-cstrt-e}), which also renders (\ref{opt-cstrt-e}) more tractable.

\subsection{COR Formulation}

To address (\ref{opt-cstrt-e}), the intractable predictive beam alignment factor $\delta(\mathbf{q}_{k,n|n-1},\mathbf{q}_{k,n})$ is approximated by its second-order Taylor expansion around $\kappa(\mathbf{q}_{k,n|n-1},\mathbf{q}_{k,n}) = 0$ given by  
\begin{equation}
    \delta(\mathbf{q}_{k,n|n-1},\mathbf{q}_{k,n}) 
    \approx 1 - \delta_{\rm{A}}\kappa(\mathbf{q}_{k,n|n-1},\mathbf{q}_{k,n})^{2}, \label{fm:aBF}
\end{equation}
where the coefficient $\delta_{\rm{A}} = \pi^{2}(N_{\rm{t}}^{2}-1)/12$ is derived from the second-order derivative.\footnote{In this paper, the case with $1 - \delta_{\rm{A}}\kappa(\mathbf{q}_{k,n|n-1},\mathbf{q}_{k,n})^{2} > 0$ is considered to assure a positive SNR.}
Given the approximated predictive beam alignment factor (\ref{fm:aBF}), the received SNR of the $k$-th UAV at the $n$-th time slot can be approximated and compactly represented by  
\vspace{-1.5mm}
\begin{equation}
    \gamma_{k,n} \approx \frac{p_{k,n}\alpha(\mathbf{q}_{k,n|n-1},\mathbf{q}_{k,n})}{ b_{k,n} }, 
    \vspace{-1.5mm}
\end{equation}
where $\alpha(\mathbf{q}_{k,n|n-1},\mathbf{q}_{k,n})$ denotes the channel quality given by 
\begin{equation}
    \alpha(\mathbf{q}_{k,n|n-1},\mathbf{q}_{k,n}) 
    \triangleq \frac{\beta_{\rm{c}}N_{\rm{t}}(1 - \delta_{\rm{A}}\kappa(\mathbf{q}_{k,n|n-1},\mathbf{q}_{k,n})^{2}) }{ N_{\rm{c}} (\|\mathbf{q}_{k,n}\|^{2} + H_{k}^{2}) }. \label{fm:alpha}
\end{equation}
Correspondingly, the achievable rate and OP of the $k$-th UAV at the $n$-th time slot are approximated by 
\vspace{-1.5mm}
\begin{align}
    &R_{k,n} \!\approx\! b_{k,n}\log_{2}\left( 1 \!+\! \frac{p_{k,n}\alpha(\mathbf{q}_{k,n|n-1},\mathbf{q}_{k,n}) }{ b_{k,n} } \right) \!=\! \tilde{R}_{k,n}, \\
    &\zeta_{k,n}(\bar{R}_{k,n}) 
    \approx \mathbb{P}(\tilde{R}_{k,n} < \bar{R}_{k,n})
    = \tilde{\zeta}_{k,n}(\bar{R}_{k,n}), \label{fm:aOP}
    \vspace{-1.5mm}
\end{align} 
respectively. 

Next, to formulate the COR, an explicit expression for (\ref{fm:aOP}) is derived from the probability distribution of the $k$-th UAV's motion state at the $n$-th time slot $\mathbf{x}_{k,n}$. 
Specifically, the ground-truth UAV position $\mathbf{q}_{k,n}$ follows the marginal distribution of $\mathbf{x}_{k,n}$, i.e., $\mathbf{q}_{k,n} \sim \bm{\mathcal{N}}(\mathbf{q}_{k,n|n-1},\bm{\Lambda}_{k,n})$, where $\bm{\Lambda}_{k,n}$ represents the position prediction MSE matrix given by
\vspace{-1.5mm}
\begin{equation}
    \bm{\Lambda}_{k,n} = 
    \begin{bmatrix}
        [\mathbf{M}_{k,n}^{\rm{p}}]_{11} & [\mathbf{M}_{k,n}^{\rm{p}}]_{13}  \\
        [\mathbf{M}_{k,n}^{\rm{p}}]_{31} & [\mathbf{M}_{k,n}^{\rm{p}}]_{33} 
    \end{bmatrix}.
    \vspace{-1.5mm}
\end{equation} 
Thus, (\ref{fm:aOP}) can be expressed as 
\vspace{-1.5mm}
\begin{equation}
    \tilde{\zeta}_{k,n}(\bar{R}_{k,n}) 
    = 1 - \int_{\mathbf{q}_{k,n}\in\mathcal{C}_{k,n}} f(\mathbf{q}_{k,n}) \mathrm{d}\mathbf{q}_{k,n}, \label{fm:cp}
    \vspace{-1.5mm}
\end{equation} 
where $f(\mathbf{q}_{k,n})$ denotes the Gaussian PDF of $\mathbf{q}_{k,n}$ and $\mathcal{C}_{k,n}$ denotes the set of all ground-truth UAV positions constituting the complementary event of signal outage, defined as 
\vspace{-1.5mm}
\begin{equation}
    \mathcal{C}_{k,n} \triangleq \{ \mathbf{q}_{k,n} \mid \tilde{R}_{k,n} \geq \bar{R}_{k,n} \}, \label{fm:COR}
    \vspace{-1.5mm}
\end{equation} 
which is referred to as the COR.
However, it is still challenging to obtain a closed-form expression of (\ref{fm:cp}) owing to the intractable channel quality $\alpha(\mathbf{q}_{k,n|n-1},\mathbf{q}_{k,n})$ in $\tilde{R}_{k,n}$. 
Therefore, we resort to require that the COR contains an uncertainty set of ground-truth UAV positions over which the integral of $f(\mathbf{q}_{k,n})$ is tractable.
By this means, the approximated OP of the $k$-th UAV (\ref{fm:aOP}) is guaranteed below its corresponding maximum tolerable OP $\epsilon_{k}$. 
Inspired by this idea, two robust optimization schemes are proposed with different uncertainty set constructions as follows.

\begin{figure}[!t]
    \centering
    \includegraphics[width=0.48\textwidth]{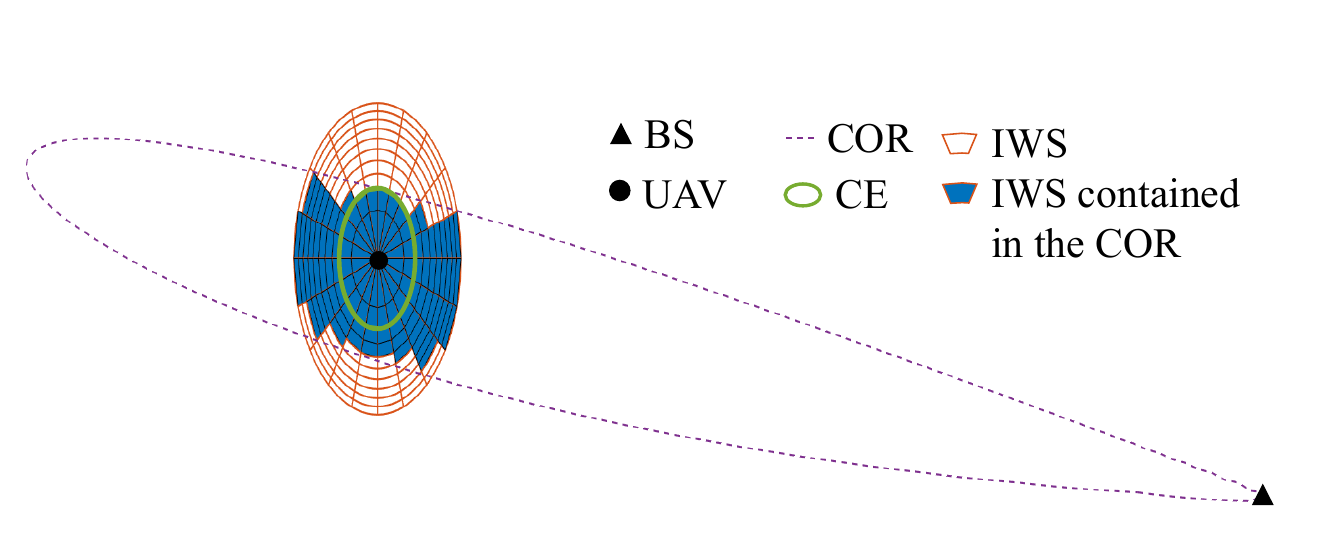}
    \caption{Illustration of proposed robust optimization schemes.} \label{fig:PA-illus}
    \vspace{-3mm}
\end{figure}

\subsection{CE-based Scheme}

Given the Gaussian distribution of $\mathbf{q}_{k,n}$, a natural choice for the uncertainty set is the CE over which the integral of $f(\mathbf{q}_{k,n})$ is equal to $1-\epsilon_{k}$.
Thus, the ground-truth UAV position $\mathbf{q}_{k,n}$ in the CE satisfies 
\vspace{-1.5mm}
\begin{equation}
    (\mathbf{q}_{k,n} - \mathbf{q}_{k,n|n-1})^{T} \bm{\Lambda}_{k,n}^{-1} (\mathbf{q}_{k,n} - \mathbf{q}_{k,n|n-1}) \leq -2\ln(\epsilon_{k}), \label{fm:CE}
    \vspace{-1.5mm}
\end{equation}
and the CE can be expressed as $\mathcal{Q}_{k,n}^{\rm{CE}} = \{ \mathbf{q}_{k,n} | \text{(\ref{fm:CE})} \}$. 
Then, by requiring $\mathcal{Q}_{k,n}^{\rm{CE}} \subseteq \mathcal{C}_{k,n}$, the maximum tolerable OP constraint (\ref{opt-cstrt-e}) is satisfied due to 
\vspace{-1.5mm}
\begin{equation}
    \tilde{\zeta}_{k,n}(\bar{R}_{k,n}) \leq 1 - \int_{\mathbf{q}_{k,n} \in\mathcal{Q}_{k,n}^{\rm{CE}} } f(\mathbf{q}_{k,n}) \mathrm{d}\mathbf{q}_{k,n} = \epsilon_{k}.
    \vspace{-1.5mm}
\end{equation}
In addition, a tight upper bound on $\bar{R}_{k,n}$ can be derived from the definition of the COR $\mathcal{C}_{k,n}$ (\ref{fm:COR}) and the relationship $\mathcal{Q}_{k,n}^{\rm{CE}} \subseteq \mathcal{C}_{k,n}$ as
\vspace{-1.5mm}
\begin{equation}
    \bar{R}_{k,n} \leq \min_{ \mathbf{q}_{k,n} \in \mathcal{Q}_{k,n}^{\rm{CE}} }\tilde{R}_{k,n}. \label{fm:tR-P2}
    \vspace{-1.5mm}
\end{equation}
Given (\ref{fm:tR-P2}), (P1) is conservatively transformed into a robust optimization problem under our proposed CE-based scheme, which is formulated as 
\vspace{-1.5mm}
\begin{equation}
    (\mathrm{P2}): \max_{ \mathbf{p}_{n}, \mathbf{b}_{n}, \{\mathbf{q}_{k,n|n-1}\}  } \min_{\mathbf{q}_{k,n} \in \mathcal{Q}_{k,n}^{\rm{CE}}, k\in\mathcal{K} } \tilde{R}_{k,n} \ 
    \text{s.t.} \ 
    \text{(\ref{opt-cstrt-a})-(\ref{opt-cstrt-d})}. \notag  
\end{equation}

\vspace{-2mm}
\subsection{IWS-based Scheme}

The aforementioned CE-based scheme only captures a local region centered at the predicted UAV position $\mathbf{q}_{k,n|n-1}$ while neglects the remaining probability distribution of the ground-truth UAV position $\mathbf{q}_{k,n}$. 
Moreover, it is still intractable to evaluate the exact OP of the $k$-th UAV at each time slot in the CE-based scheme. 
To overcome these limitations, we propose to construct the uncertainty set from IWSs detailed as follows. 

Inspired by the CE-based scheme, the IWSs corresponding to the $k$-th UAV are constructed from a sufficiently large CE centered at $\mathbf{q}_{k,n|n-1}$ to approximate the probability distribution of $\mathbf{q}_{k,n}$ with controlled truncation error, as illustrated in Fig. \ref{fig:PA-illus}. 
To be specific, $\mathbf{q}_{k,n}$ in the considered CE satisfies 
\vspace{-1.5mm}
\begin{equation}
    (\mathbf{q}_{k,n} - \mathbf{q}_{k,n|n-1})^{T} \bm{\Lambda}_{k,n}^{-1} (\mathbf{q}_{k,n} - \mathbf{q}_{k,n|n-1}) \leq -2\ln(\epsilon_{\rm{s}}), \label{fm:CE-s}
    \vspace{-1.5mm}
\end{equation}
where $\epsilon_{\rm{s}} \ll \min_{k}\epsilon_{k}$ denotes a small truncation error, and thus the considered CE can be denoted by $\mathcal{Q}_{k,n}^{\rm{s}} = \{ \mathbf{q}_{k,n} | \text{(\ref{fm:CE-s})} \}$. 
Then, a whitening transform $U(\cdot)$ is applied to the CE $\mathcal{Q}_{k,n}^{\rm{s}}$, and the whitened UAV position prediction error vector $\mathbf{q}^{\rm{s}} \sim \bm{\mathcal{N}}(\mathbf{0},\mathbf{I}_{2})$ is expressed as \cite{whitening0}
\vspace{-1.5mm}
\begin{equation}
    \mathbf{q}^{\rm{s}} = U(\mathbf{q}_{k,n};\mathbf{q}_{k,n|n-1}) 
    = \bm{\Lambda}_{k,n}^{-\frac{1}{2}}(\mathbf{q}_{k,n} - \mathbf{q}_{k,n|n-1}). \label{fm:U}
    \vspace{-1.5mm}
\end{equation}
Accordingly, the whitened CE is denoted by $\mathcal{Q}^{\rm{s}} = \{\mathbf{q}^{\rm{s}} \mid \text{(\ref{fm:U})}, \forall \mathbf{q}_{k,n} \in \mathcal{Q}_{k,n}^{\rm{s}} \}$. 
Note that both $\mathcal{Q}^{\rm{s}}$ and $\mathbf{q}^{\rm{s}}$ are independent of the subscripts $k$ and $n$ after the whitening transform. 
Next, the whitened CE $\mathcal{Q}^{\rm{s}}$ is uniformly discretized into $I_{\rm{g}}=I_{\rm{r}}I_{\rm{a}}$ sectors along the radial and angular dimensions, where $I_{\rm{r}}$ and $I_{\rm{a}}$ denote the numbers of radial and angular intervals, respectively. 
The radial and angular intervals are given by 
\vspace{-1.5mm}
\begin{equation}
    \Delta r^{\rm{s}} = \frac{\sqrt{-2\ln(\epsilon_{\rm{s}})}}{I_{\rm{r}}}, \quad
    \Delta \theta^{\rm{s}} = \frac{2\pi}{I_{\rm{a}}}. \label{fm:dr}
    \vspace{-1.5mm}
\end{equation}
Meanwhile, the discretized sectors are indexed by $(i,j)$ and the corresponding index set is denoted by 
\vspace{-1.5mm}
\begin{equation}
    \mathcal{L}^{\rm{s}} = \{ (i,j) | i=1,\ldots,I_{\rm{r}}, j=1,\ldots,I_{\rm{a}} \}. \label{fm:Lij}
    \vspace{-1.5mm}
\end{equation}
Given (\ref{fm:dr}) and (\ref{fm:Lij}), the $(i,j)$-th sector is expressed as 
\vspace{-1.5mm}
\begin{align}
    \!\!\!\!\!\!\mathcal{S}_{(i,j)} 
    &= \{ \mathbf{q}^{\rm{s}} | \mathbf{q}^{\rm{s}} 
    = [r^{\rm{s}}\cos{(\theta^{\rm{s}})}, r^{\rm{s}}\sin{(\theta^{\rm{s}})}]^{T}, \notag \\
    \underline{r}_{i}^{\rm{s}} 
    &= (i-1)\Delta r^{\rm{s}} \leq r^{\rm{s}} \leq i \Delta r^{\rm{s}} = \overline{r}_{i}^{\rm{s}}, \notag \\ 
    \underline{\theta}_{j}^{\rm{s}} 
    &= (j-1)\Delta \theta^{\rm{s}} \leq \theta^{\rm{s}} \leq j\Delta \theta^{\rm{s}} = \overline{\theta}_{j}^{\rm{s}}, (i,j) \in \mathcal{L}^{\rm{s}} \}, 
    \vspace{-1.5mm}
\end{align}
where $\underline{r}_{i}^{\rm{s}}$ and $\overline{r}_{i}^{\rm{s}}$ denote the minimum and maximum range of the $(i,j)$-th sector in the whitened domain, respectively, while $\underline{\theta}_{j}^{\rm{s}}$ and $\overline{\theta}_{j}^{\rm{s}}$ denote the minimum and maximum azimuth angle of the $(i,j)$-th sector, respectively. 
Correspondingly, the set of corner points enclosing the $(i,j)$-th sector is expressed as 
\vspace{-1.5mm}
\begin{equation}
    \mathcal{G}_{(i,j)}^{\rm{s}} = \{ \mathbf{q}^{\rm{s}} | 
    r^{\rm{s}} \in \{ \underline{r}_{i}^{\rm{s}}, \overline{r}_{i}^{\rm{s}} \}, \theta^{\rm{s}} \in \{ \underline{\theta}_{j}^{\rm{s}}, \overline{\theta}_{j}^{\rm{s}} \}, (i,j)\in\mathcal{L}^{\rm{s}} \}. 
    \vspace{-1.5mm}
\end{equation}
Finally, IWSs are obtained by applying the inverse whitening transform $U^{-1}(\cdot)$ to the sectors $\mathcal{S}_{(i,j)}, \forall (i,j) \in \mathcal{L}^{\rm{s}}$. 
For notational simplicity, IWSs are indexed by $(i,j)$ as well.
Particularly, the $(i,j)$-th IWS and its corner point set are represented by 
\vspace{-1.5mm}
\begin{align}
    \!\!\mathcal{S}_{k,n,(i,j)}^{\text{IW}} 
    &\!\!=\! \{ \mathbf{q}_{k,n} | \mathbf{q}_{k,n} \!=\! U^{-1}(\mathbf{q}^{\rm{s}};\mathbf{q}_{k,n|n-1}), \mathbf{q}^{\rm{s}} \!\in\! \mathcal{S}_{(i,j)} \}, \label{fm:S-iws} \\
    \!\!\mathcal{G}_{k,n,(i,j)}^{\text{IW}} 
    &\!\!=\! \{ \mathbf{q}_{k,n} | \mathbf{q}_{k,n} \!=\! U^{-1}(\mathbf{q}^{\rm{s}};\mathbf{q}_{k,n|n-1}), \mathbf{q}^{\rm{s}} \!\in\! \mathcal{G}_{(i,j)}^{\rm{s}} \}, \label{fm:G-iws}
    \vspace{-1.5mm}
\end{align}
respectively. 

Based on IWSs, the OP of the $k$-th UAV at the $n$-th time slot is approximated as follows. 
First, given the outage capacity $\bar{R}_{k,n}$, the $(i,j)$-th IWS is identified as contained in the COR if it satisfies\footnote{Although such identification is in fact an approximation, its effectiveness is validated by the OP approximation accuracy in Section V-A.} 
\vspace{-1.5mm}
\begin{equation}
    \min_{\mathbf{q}_{k,n} \in \mathcal{G}_{k,n,(i,j)}^{\text{IW}}}  \tilde{R}_{k,n}  > \bar{R}_{k,n}. \label{fm:s-cout}
    \vspace{-1.5mm}
\end{equation}
Thus, the overall corner point set of IWSs contained in the COR can be expressed as 
\vspace{-1.5mm}
\begin{equation}
    \mathcal{G}_{k,n}^{\text{IW,co}} = \bigcup_{(i,j)\in \mathcal{L}_{k,n}^{\rm{co}}}\mathcal{G}_{k,n,(i,j)}^{\text{IW}}, \label{fm:s-iw-co}
    \vspace{-1.5mm}
\end{equation}
where $\mathcal{L}_{k,n}^{\rm{co}} = \{(i,j)|\text{(\ref{fm:s-cout})}, (i,j)\in\mathcal{L}^{\rm{s}}\}$ denotes the index set of IWSs identified as contained in the COR. 
Then, the complementary OP associated with the $(i,j)$-th IWS, defined by the integral of the ground-truth UAV position PDF $f(\mathbf{q}_{k,n})$ over the $(i,j)$-th IWS, is derived as\footnote{Although the index $j$ does not explicitly appear in the expression of $\zeta_{(i,j)}^{\rm{s}}$ according to (\ref{fm:zeta}), it is retained in the subscript to indicate the association of $\zeta_{(i,j)}^{\rm{s}}$ with the $(i,j)$-th IWS.} 
\vspace{-1.5mm}
\begin{align}
    \zeta_{(i,j)}^{\rm{s}} 
    &= \int_{\mathbf{q}_{k,n} \in \mathcal{S}_{k,n,(i,j)}^{\text{IW}} } f(\mathbf{q}_{k,n}) \mathrm{d}\mathbf{q}_{k,n} = \int_{\mathbf{q}^{\rm{s}} \in \mathcal{S}_{(i,j)} } f(\mathbf{q}^{\rm{s}}) \mathrm{d}\mathbf{q}^{\rm{s}} \notag \\
    &=\frac{\Delta \theta^{\rm{s}}}{2\pi} \left( \mathrm{exp}\Bigl(-\frac{1}{2}\Bigl(\underline{r}_{i}^{\rm{s}}\Bigr)^{2}\Bigr) - \mathrm{exp}\Bigl(-\frac{1}{2} \Bigl(\overline{r}_{i}^{\rm{s}} \Bigr)^{2} \Bigr) \right), \label{fm:zeta}
    \vspace{-1.5mm}
\end{align}
where $f(\mathbf{q}^{\rm{s}})$ denotes the PDF of $\mathbf{q}^{\rm{s}}$.
Given $\mathcal{L}_{k,n}^{\rm{co}}$ and (\ref{fm:zeta}), the OP (\ref{fm:cp}) can be approximated based on the sum of all complementary OPs associated with the IWSs contained in the COR, i.e., 
\vspace{-1.5mm}
\begin{equation}
    \tilde{\zeta}_{k,n}(\bar{R}_{k,n}) 
    \approx \tilde{\zeta}_{k,n}^{\text{IW}}(\bar{R}_{k,n}) 
    = 1 - \sum_{(i,j)\in \mathcal{L}_{k,n}^{\rm{co}} } \zeta_{(i,j)}^{\rm{s}}, \label{fm:aOP-SG}
    \vspace{-1.5mm}
\end{equation}
where a high approximation accuracy can be achieved by sufficiently small $\Delta r^{\rm{s}}$ and $\Delta \theta^{\rm{s}}$.
Therefore, in our proposed IWS-based scheme, (P1) is approximated by a robust optimization problem formulated as 
\vspace{-1.5mm}
\begin{align}
    (\mathrm{P3}): \ &\max_{ \mathbf{p}_{n}, \mathbf{b}_{n}, \{\mathbf{q}_{k,n|n-1}\}  } \min_{\mathbf{q}_{k,n} \in \mathcal{G}_{k,n}^{\text{IW,co}}, k\in\mathcal{K}} \tilde{R}_{k,n}  \label{opt3} \\ 
    \text{s.t.} \ 
    &1 - \sum_{(i,j)\in \mathcal{L}_{k,n}^{\rm{co}} } \zeta_{(i,j)}^{\rm{s}} < \epsilon_{k}, \forall k, \tag{\ref{opt3}{a}} \label{opt3-a} \\
    &\text{(\ref{opt-cstrt-a})-(\ref{opt-cstrt-d})}, \notag
    \vspace{-1.5mm}
\end{align} 
where the objective function is derived from (\ref{fm:s-cout}) and the constraint (\ref{opt3-a}) approximates the maximum tolerable OP constraint (\ref{opt-cstrt-e}) in (P1).

\section{Proposed Algorithms and A Unified Optimization Framework}

In this section, an efficient algorithm for (P2) is proposed, where the predicted UAV positions $\{\mathbf{q}_{k,n|n-1}\}$ and resource allocation $\{\mathbf{p}_{n},\mathbf{b}_{n}\}$ are sequentially optimized.
Inspired by the algorithm for (P2), a sorting-based selection of IWSs is proposed to decouple $\{\mathbf{q}_{k,n|n-1}\}$ and $\{\mathbf{p}_{n},\mathbf{b}_{n}\}$ in the constraint (\ref{opt3-a}). 
Based on the proposed IWS selection, an efficient algorithm for (P3) is proposed. 
Furthermore, the algorithms for (P2) and (P3) are summarized into a unified framework due to their similarities to obtain insights into robust UAV trajectory and resource optimization. 

\subsection{Algorithm for (P2)}

Note that the predicted UAV positions $\{\mathbf{q}_{k,n|n-1}\}$ are irrelevant to the power and bandwidth budget constraints (\ref{opt-cstrt-a})-(\ref{opt-cstrt-c}) in (P2), and the CEs among different UAVs are independent. 
Moreover, the approximated achievable rate $\tilde{R}_{k,n}$ is a monotonically increasing function of the channel quality $\alpha(\mathbf{q}_{k,n|n-1},\mathbf{q}_{k,n})$ given any $p_{k,n}$ and $b_{k,n}$. 
Therefore, the predicted UAV positions $\{\mathbf{q}_{k,n|n-1}\}$ can be optimized by individually solving a subproblem of (P2) for each UAV, which is formulated as 
\vspace{-1.5mm}
\begin{align}
    (\mathrm{P2'}): \ &\max_{ \mathbf{q}_{k,n|n-1} } \ \ \min_{\acute{\mathbf{q}}_{k,n}} \alpha(\mathbf{q}_{k,n|n-1},\mathbf{q}_{k,n|n-1} + \acute{\mathbf{q}}_{k,n}) \label{opt-2sl}  \\ 
    \text{s.t.} \ 
    &\|\mathbf{q}_{k,n|n-1} - \hat{\mathbf{q}}_{k,n-1} \| \leq V_{\rm{max}}\Delta T, \tag{\ref{opt-2sl}{a}} \label{opt-2sl-a} \\
    &(\acute{\mathbf{q}}_{k,n})^{T} \bm{\Lambda}_{k,n}^{-1} \acute{\mathbf{q}}_{k,n} + 2\ln(\epsilon_{k}) \leq 0, \tag{\ref{opt-2sl}{b}} \label{opt-2sl-b}
    \vspace{-1.5mm}
\end{align}
where $\acute{\mathbf{q}}_{k,n} = \mathbf{q}_{k,n} - \mathbf{q}_{k,n|n-1}$ is an auxiliary variable to decouple $\mathbf{q}_{k,n|n-1}$ from $\mathbf{q}_{k,n}$. 
In fact, ($\mathrm{P2'}$) is also equivalent to a predicted UAV position optimization problem in the single-UAV case with $K=1$. 

Thanks to the decoupled optimization variables $\mathbf{q}_{k,n|n-1}$ and $\acute{\mathbf{q}}_{k,n}$ in ($\mathrm{P2'}$), the non-convex objective function in ($\mathrm{P2'}$) can be addressed by the BSCA technique to obtain a locally optimal solution, where $\mathbf{q}_{k,n|n-1}$ and $\acute{\mathbf{q}}_{k,n}$ are optimized by solving two surrogate optimization problems in an alternating manner \cite{Raza,BSCA-UAV}.
Specifically, let $\tilde{\mathbf{q}}_{1,m}$ and $\tilde{\mathbf{q}}_{2,m}$ denote the predicted UAV position and auxiliary variable to be optimized in the $m$-th iteration, respectively, where the subscripts $k$ and $n$ are ignored for notational simplicity. 
Meanwhile, $\tilde{\mathbf{Q}}_{m-1}^{*} \triangleq \{ \tilde{\mathbf{q}}_{1,m-1}^{*}, \tilde{\mathbf{q}}_{2,m-1}^{*} \}$ is defined as the set of solutions obtained in the $(m-1)$-th iteration of the BSCA algorithm.
Then, in the $m$-th iteration, the $w$-th block of optimization variables $\tilde{\mathbf{q}}_{w,m}$ is updated with $w= (m\ \mathrm{mod}\ 2)+1$ by solving the subproblem ($\mathrm{P2'}.w$) formulated as 
\vspace{-1.5mm}
\begin{equation}
    \left\{ \begin{aligned}
        &(\mathrm{P2'}.1): \ \max_{ \tilde{\mathbf{q}}_{1,m} } \alpha_{1}(\tilde{\mathbf{q}}_{1,m};\tilde{\mathbf{Q}}_{m-1}^{*}) \
    \text{s.t.} \ \text{(\ref{opt-2sl-a})}, &w=1, \notag \\
        &(\mathrm{P2'}.2): \ \min_{ \tilde{\mathbf{q}}_{2,m} } \alpha_{2}(\tilde{\mathbf{q}}_{2,m};\tilde{\mathbf{Q}}_{m-1}^{*}) \ 
    \text{s.t.} \ \text{(\ref{opt-2sl-b})}, &w=2, \notag
    \end{aligned} \right. \notag
    \vspace{-1.5mm}
\end{equation}
where $\alpha_{1}(\cdot)$ and $\alpha_{2}(\cdot)$ denote the surrogate objective functions approximating the function $\alpha(\cdot)$ expressed as 
\vspace{-1.5mm}
\begin{align}
    &\alpha_{w}(\tilde{\mathbf{q}}_{w,m};\tilde{\mathbf{Q}}_{m-1}^{*}) = \alpha( \tilde{\mathbf{q}}_{1,m-1}^{*}, \tilde{\mathbf{q}}_{1,m-1}^{*} + \tilde{\mathbf{q}}_{2,m-1}^{*} ) \notag \\
    &+ \nabla_{\tilde{\mathbf{q}}_{w,m}} \alpha( \tilde{\mathbf{q}}_{1,m-1}^{*}, \tilde{\mathbf{q}}_{1,m-1}^{*} + \tilde{\mathbf{q}}_{2,m-1}^{*} )^{T}(\tilde{\mathbf{q}}_{w,m} - \tilde{\mathbf{q}}_{w,m-1}^{*}) \notag \\
    &+ \frac{1}{2}(\tilde{\mathbf{q}}_{w,m} \!-\! \tilde{\mathbf{q}}_{w,m-1}^{*})\mathbf{Q}_{w}(\tilde{\mathbf{q}}_{w,m} \!-\! \tilde{\mathbf{q}}_{w,m-1}^{*})^{T}, w=1,2, \label{fm:BSCA}
    \vspace{-1.5mm}
\end{align}
and $\mathbf{Q}_{w}$ denotes a given matrix ensuring the convexity of ($\mathrm{P2'}.w$) \cite{BSCA-UAV}.

Given the obtained solution to ($\mathrm{P2'}$) denoted by $\{\mathbf{q}_{k,n|n-1}^{*}, \acute{\mathbf{q}}_{k,n}^{*}\}$, the worst-case channel quality when the predicted UAV positions are designed as $\{\mathbf{q}_{k,n|n-1}^{*}\}$ in the CE-based scheme is determined by 
\vspace{-1.5mm}
\begin{equation}
    \alpha_{k,n}^{*} = \alpha(\mathbf{q}_{k,n|n-1}^{*},\acute{\mathbf{q}}_{k,n}^{*} + \mathbf{q}_{k,n|n-1}^{*}). \label{fm:wcq-CE}
    \vspace{-1.5mm}
\end{equation}
By substituting (\ref{fm:wcq-CE}) into $\tilde{R}_{k,n}$, (P2) reduces to a subproblem for joint power and bandwidth allocation among $K$ UAVs formulated as 
\vspace{-1.5mm}
\begin{equation}
    (\mathrm{P2''}): \ \max_{ \mathbf{p}_{n}, \mathbf{b}_{n} } \ \min_{k\in\mathcal{K}}  b_{k,n}\log_{2}\Bigl( 1 + \frac{p_{k,n}\alpha_{k,n}^{*} }{ b_{k,n} } \Bigr)  \ 
    \text{s.t.} \ 
    \text{(\ref{opt-cstrt-a})-(\ref{opt-cstrt-c})}, \notag 
    \vspace{-1.5mm}
\end{equation}
which is a classic convex \emph{max-min} problem and can be optimally solved by standard approaches such as the interior-point method \cite{potra2000interior}.

\subsection{Algorithm for (P3)}

Different from (P2), the optimization variables in (P3) are coupled in the overall corner point set of IWSs contained in the COR $\mathcal{G}_{k,n}^{\text{IW,co}}$ and its corresponding index set $\mathcal{L}_{k,n}^{\rm{co}}$ due to the identification (\ref{fm:s-cout}). 
However, if the IWSs contained in the COR are determined by a given selection satisfying both (\ref{fm:s-cout}) and the approximated maximum tolerable OP constraint (\ref{opt3-a}), then the sets $\mathcal{G}_{k,n}^{\text{IW,co}}$, $\mathcal{L}_{k,n}^{\rm{co}}$ no longer rely on (\ref{fm:s-cout}) and the optimization variables in (P3) can be decoupled. 
Therefore, such an IWS selection is proposed based on the channel quality evaluated at the corner points of IWSs. 
Specifically, define the worst-case channel quality associated with the $(i,j)$-th IWS of the $k$-th UAV at the $n$-th time slot by 
\vspace{-1.5mm}
\begin{equation}
    \underline{\alpha}_{(i,j)}(\mathbf{q}_{k,n|n-1}) \triangleq \min_{\mathbf{q}_{k,n} \in \mathcal{G}_{k,n,(i,j)}^{\text{IW}}} \alpha(\mathbf{q}_{k,n|n-1}, \mathbf{q}_{k,n}), \label{fm:wc-q-ij}
    \vspace{-1.5mm}
\end{equation}
which represents the minimum channel qualities evaluated at the $(i,j)$-th IWS's four corner points given $\mathbf{q}_{k,n|n-1}$. 
Then, all IWSs of the $k$-th UAV are sorted in ascending order of their associated worst-case channel qualities with the order index denoted by $\iota_{k,n}=1,...,I_{\rm{g}}$.
Let $S(\iota_{k,n})=(i,j)$ denote the mapping from the order index $\iota_{k,n}$ to the IWS index $(i,j)$.
Based on $\iota_{k,n}$, we propose to select the IWSs with $\iota_{k,n}\geq\iota_{k,n}^{*}$, where $\iota_{k,n}^{*}$ denotes the minimum order index satisfying
\vspace{-1.5mm}
\begin{equation}
    \sum_{\iota_{k,n}=\iota_{k,n}^{*}}^{I_{\rm{g}}} \zeta_{S(\iota_{k,n})}^{\rm{s}} > 1-\epsilon_{k}. \label{fm:sel}
    \vspace{-1.5mm}
\end{equation}
With (\ref{fm:sel}), the constraint (\ref{opt3-a}) is satisfied. 
As such, the index set of selected IWSs is denoted by
\vspace{-1.5mm}
\begin{equation}
    \mathcal{L}_{k,n}^{*}(\mathbf{q}_{k,n|n-1}) = \{S(\iota_{k,n})| \iota_{k,n} = \iota_{k,n}^{*}, \ldots, I_{\rm{g}} \}, \label{fm:L-sel}
    \vspace{-1.5mm}
\end{equation}
and the overall corner point set of selected IWSs can be expressed as 
\vspace{-1.5mm}
\begin{equation}
    \mathcal{G}_{k,n}^{\text{IW},*} = \bigcup_{(i,j)\in \mathcal{L}_{k,n}^{*}(\mathbf{q}_{k,n|n-1}) }\mathcal{G}_{k,n,(i,j)}^{\text{IW}}. \label{fm:G-tot}
    \vspace{-1.5mm}
\end{equation}
Given the selected IWSs, the overall corner point set of IWSs contained in the COR $\mathcal{G}_{k,n}^{\text{IW,co}}$ and its corresponding index set $\mathcal{L}_{k,n}^{\rm{co}}$ in (P3) are replaced with $\mathcal{G}_{k,n}^{\text{IW},*}$ and $\mathcal{L}_{k,n}^{*}(\mathbf{q}_{k,n|n-1})$, respectively, and the resulting robust optimization problem is formulated as 
\vspace{-1.5mm}
\begin{equation}
    (\mathrm{P4}): \ \max_{ \mathbf{p}_{n}, \mathbf{b}_{n}, \{\mathbf{q}_{k,n|n-1}\}  } \min_{\mathbf{q}_{k,n} \in \mathcal{G}_{k,n}^{\text{IW},*}, k\in\mathcal{K}} \tilde{R}_{k,n}  \
    \text{s.t.} \ 
    \text{(\ref{opt-cstrt-a})-(\ref{opt-cstrt-d})}, \notag
    \vspace{-1.5mm}
\end{equation} 
where the predicted UAV positions $\{\mathbf{q}_{k,n|n-1}\}$ are decoupled from $\{\mathbf{p}_{n},\mathbf{b}_{n}\}$ because the index set $\mathcal{L}_{k,n}^{*}(\mathbf{q}_{k,n|n-1})$ is obtained regardless of $\mathbf{p}_{n}$ and $\mathbf{b}_{n}$.

\begin{remark}
    To solve (P3) straightforwardly, a common approach is to introduce Boolean variables denoting the IWS selection and formulate (P3) as a mixed integer non-linear program (MINLP), which is usually solved by the branch-and-bound or big-M method \cite{RA-tutor,burer2012non}. 
    However, the computational complexities of these approaches grow rapidly (exponentially in the worst case) with the total number of IWSs $I_{\rm{g}}$ and thus can be unacceptably high resulting from small $\Delta r^{\rm{s}}$ and $\Delta \theta^{\rm{s}}$.
    In comparison, the computational complexity of our proposed IWS selection scales only near-linearly with $I_{\rm g}$ resulting from the sorting, which is more suitable for real-time trajectory and resource optimization.
\end{remark}

Note that (P4) and (P2) share highly similar formulations with the only difference lying in the uncertainty set of $\mathbf{q}_{k,n}$. 
Therefore, similar to the algorithm for (P2), the predicted UAV positions $\{\mathbf{q}_{k,n|n-1}\}$ are optimized individually for each UAV, and the optimization problem for the $k$-th UAV can be compactly formulated as 
\vspace{-1.5mm}
\begin{equation}
    (\mathrm{P4'}): \ \max_{ \mathbf{q}_{k,n|n-1} } \ \underline{\alpha}_{S(\iota_{k,n}^{*})}(\mathbf{q}_{k,n|n-1}) \  
    \text{s.t.} \ 
    \text{(\ref{opt-2sl-a})}. \notag 
    \vspace{-1.5mm}
\end{equation}
Particularly, the objective function in ($\mathrm{P4'}$) is derived as 
\vspace{-1.5mm}
\begin{align}
    &\min_{\mathbf{q}_{k,n} \in \mathcal{G}_{k,n}^{\text{IW},*}} \alpha(\mathbf{q}_{k,n|n-1},\mathbf{q}_{k,n}) \notag \\
    &\overset{(a)}{=} \min_{(i,j)\in \mathcal{L}_{k,n}^{*}(\mathbf{q}_{k,n|n-1})} \min_{ \mathbf{q}_{k,n} \in \mathcal{G}_{k,n,(i,j)}^{\text{IW}} } \alpha(\mathbf{q}_{k,n|n-1},\mathbf{q}_{k,n}) \notag \\
    &\overset{(b)}{=} \min_{(i,j)\in \mathcal{L}_{k,n}^{*}(\mathbf{q}_{k,n|n-1})} \underline{\alpha}_{(i,j)}(\mathbf{q}_{k,n|n-1}) \notag \\
    &\overset{(c)}{=} \underline{\alpha}_{S(\iota_{k,n}^{*})}(\mathbf{q}_{k,n|n-1}), \label{fm:deriv}
    \vspace{-1.5mm}
\end{align}
where $(a)$, $(b)$, and $(c)$ are due to (\ref{fm:G-tot}), (\ref{fm:wc-q-ij}), and the ascending-order sorting in our proposed IWS selection, respectively. 
The derivation (\ref{fm:deriv}) shows that resulting from our proposed IWS selection, $\mathbf{q}_{k,n}$ is determined by the corner point leading to the worst-case channel quality associated with the $S(\iota_{k,n}^{*})$-th IWS given $\mathbf{q}_{k,n|n-1}$, expressed as 
\vspace{-1.5mm}
\begin{equation}
    \mathbf{q}_{k,n} = \argmin_{\mathbf{q}_{k,n}\in \mathcal{G}_{k,n,S(\iota_{k,n}^{*})}^{\text{IW}}} \alpha(\mathbf{q}_{k,n|n-1},\mathbf{q}_{k,n}). 
    \vspace{-1.5mm}
\end{equation}
Thus, $\mathbf{q}_{k,n|n-1}$ is the only optimization variable in ($\mathrm{P4'}$) and the PGD approach \cite{PGD} can be applied to obtain a locally optimal solution based on an approximated gradient given by $\tilde{\nabla}_{\mathbf{q}_{k,n|n-1}} \underline{\alpha}_{S(\iota_{k,n}^{*})}(\mathbf{q}_{k,n|n-1}) \in\mathbb{R}^{2\times 1}$.
The $i_{\rm{d}}$-th component of the approximated gradient is denoted by 
\vspace{-1.5mm}
\begin{align}
    &[ \tilde{\nabla}_{\mathbf{q}_{k,n|n-1}} \underline{\alpha}_{S(\iota_{k,n}^{*})}(\mathbf{q}_{k,n|n-1}) ]_{i_{\rm{d}}} \notag \\
    &=
    \frac{ \underline{\alpha}_{S(\iota_{k,n}^{*})}(\mathbf{q}_{k,n|n-1} + \Delta_{i_{\rm{d}}}\bm{e}_{i_{\rm{d}}}) - \underline{\alpha}_{S(\iota_{k,n}^{*})}(\mathbf{q}_{k,n|n-1})}{\Delta_{i_{\rm{d}}}}, \label{fm:aGrad}
    \vspace{-1.5mm}
\end{align}
where $\Delta_{i_{\rm{d}}}$ denotes the approximated differential length of the $i_{\rm{d}}$-th component and $\bm{e}_{i_{\rm{d}}}$ denotes the unit vector with its $i_{\rm{d}}$-th component equal to 1 with $i_{\rm{d}}=1,2$.
Particularly, let $\mathbf{q}_{q}$ denote the optimization variable $\mathbf{q}_{k,n|n-1}$ in the $q$-th iteration.
Subsequently, the original updated $\mathbf{q}_{q}$ is given by 
\vspace{-1.5mm}
\begin{equation}
    \tilde{\mathbf{q}}_{q+1} = \mathbf{q}_{q} - \nu_{q} \tilde{\nabla}_{\mathbf{q}_{k,n|n-1}} \underline{\alpha}_{S(\iota_{k,n}^{*})}(\mathbf{q}_{q}),
    \vspace{-1.5mm}
\end{equation}
and the final update is expressed as 
\vspace{-1.5mm}
\begin{equation}
    \mathbf{q}_{q+1} 
    \!=\! \hat{\mathbf{q}}_{n-1} \!+\! \min\Bigl\{1,\frac{V_{\rm{max}}\Delta T}{\|\tilde{\mathbf{q}}_{q+1} \!-\! \hat{\mathbf{q}}_{n-1}\|}\Bigr\}(\tilde{\mathbf{q}}_{q+1} \!-\! \hat{\mathbf{q}}_{n-1}),
    \vspace{-1.5mm}
\end{equation}
where $\nu_{q}$ denotes the step size in the $q$-th iteration. 
After solving ($\mathrm{P4'}$), the worst-case channel quality given the optimized predicted UAV positions $\{\mathbf{q}_{k,n|n-1}^{*}\}$ in the IWS-based scheme is determined by 
\vspace{-1.5mm}
\begin{equation}
    \alpha_{k,n}^{*} = \underline{\alpha}_{S(\iota_{k,n}^{*})}(\mathbf{q}_{k,n|n-1}^{*}). \label{fm:wcq-IWS}
    \vspace{-1.5mm}
\end{equation}
Given $\alpha_{k,n}^{*}$, the resulting subproblem of (P4) for joint power and bandwidth allocation is exactly the same as ($\mathrm{P2''}$).

\subsection{Unified Framework and Complexity Analysis}

By summarizing the proposed algorithms for (P2) and (P3), a unified framework is proposed for robust UAV trajectory and resource optimization, where the individual predicted UAV position and joint resource allocation are sequentially optimized by solving the following two optimization problems formulated as
\vspace{-1.5mm}
\begin{align}
    &\max_{ \mathbf{q}_{k,n|n-1} } \ \min_{\mathbf{q}_{k,n}\in\mathcal{X}} \alpha(\mathbf{q}_{k,n|n-1},\mathbf{q}_{k,n}) \ \notag \\
    \text{s.t.} \ 
    &\text{Position-related constraints}, \notag \\
    &\max_{ \mathbf{p}_{n}, \mathbf{b}_{n} } \ \min_{k\in\mathcal{K}}  b_{k,n}\log_{2}\Bigl( 1 + \frac{p_{k,n}\alpha_{k,n}^{*} }{ b_{k,n} } \Bigr) \ \notag \\
    \text{s.t.} \ 
    &\text{Resource-related constraints}, \notag 
    \vspace{-1.5mm}
\end{align}
where $\mathcal{X}\in\{ \mathcal{Q}_{k,n}^{\rm{CE}}, \mathcal{G}_{k,n}^{\text{IW},*} \}$ denotes the constructed uncertainty set according to the specific proposed robust optimization schemes, and $\alpha_{k,n}^{*}$ denotes the maximized objective value of the first optimization problem. 
Such a sequential optimization structure avoids the multiple iterations of alternating optimization and thus is of low computational complexity. 
Note that the IWSs obtained by our proposed sorting-based selection can also be considered as an uncertainty set with some specific property. 
Therefore, the key to applying the unified framework is to construct an appropriate uncertainty set to decouple the optimization variables while ensuring solution robustness.
Moreover, our work shows that the proposed framework applies to both continuous and discrete uncertainty sets, i.e., the CE and IWS, respectively.

Furthermore, the computational complexities of the proposed algorithms for (P2) and (P3) are analyzed as follows. 
In the CE-based scheme, since the dimension of each optimization block is 2 in ($\mathrm{P2'}$), the main computational complexity of the algorithm for ($\mathrm{P2'}$) is given by $\mathcal{O}(2^{3.5}I_{\rm{BSCA}})$, where $I_{\rm{BSCA}}$ denotes the number of iterations required for the convergence of BSCA \cite{BSCA-UAV}. 
While in the IWS-based scheme, the main computational complexity of solving ($\mathrm{P4'}$) is given by $\mathcal{O}(I_{\rm{PGD}}I_{\rm{g}}\log I_{\rm{g}})$, where $I_{\rm{PGD}}$ denotes the number of iterations required for the convergence of PGD and $\mathcal{O}(I_{\rm{g}}\log I_{\rm{g}})$ is due to the computational complexity needed for the sorting of IWSs. 
Besides, the computational complexity of solving ($\mathrm{P2''}$) by the interior-point method is given by $\mathcal{O}(K^{3.5})$ \cite{potra2000interior}. 
Thus, considering the total number of UAVs $K$, the main computational complexities of the proposed algorithms are given by $\mathcal{O}(2^{3.5}I_{\rm{BSCA}}K+K^{3.5})$ and $\mathcal{O}(KI_{\rm{PGD}}I_{\rm{g}}\log I_{\rm{g}}+K^{3.5})$, respectively.

\begin{figure}[!t]
    \centering
    \includegraphics[width=0.38\textwidth]{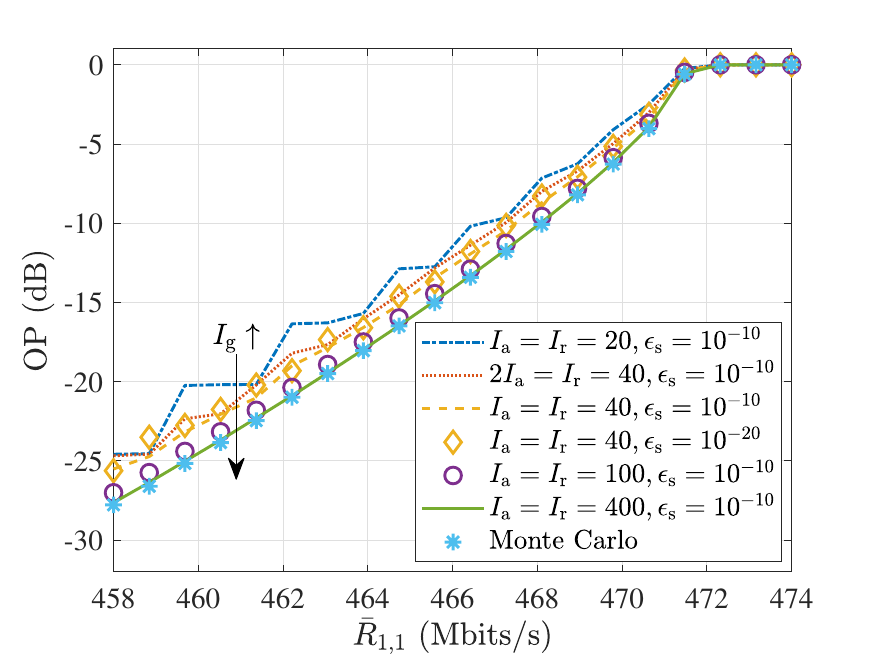}
    \caption{The IWS-based OP approximation accuracy verifications.}
    \label{fig:A-SG}
    \vspace{-3mm}
\end{figure}

\section{Numerical Results}

In this section, numerical results are provided to illustrate the effectiveness and performance gain of our proposed robust optimization schemes. 
Unless specified otherwise, the following system parameters are used: $V_{\rm{max}} = 30$ m/s, $\Delta T = 0.02$ s, $N_{\rm{c}} = -174$ dBm/Hz, $\lambda = 0.01$ m, $\sigma_{\rm{rcs}} = 0.2 \text{m}^{2}$, $N_{\rm{t}} = N_{\rm{r}} = 64$, $\tilde{q} = 0.1$, $a^{\rm{dst}} = 9$, $G_{\rm{ar}} = N_{\rm{t}}N_{\rm{r}} = 4096$, $G_{\rm{mf}} = 5\times 10^{3}$ \cite{YFJ2024CL,YFJ2026TWC,KTMeng2024TVT-IS}.

\subsection{OP Approximation and Algorithms Verification} 

Fig. \ref{fig:A-SG} illustrates the accuracy of the proposed IWS-based OP approximation (\ref{fm:aOP-SG}) with different $I_{\rm{a}}$ and $I_{\rm{r}}$.
Specifically, at a typical time slot $n=1$, the random ground-truth position $\mathbf{q}_{1,1}$ of a single UAV with $k=1$ is simulated via $10^{6}$ Monte Carlo trials. 
The predicted UAV position and the position prediction MSE matrix are set as $\mathbf{q}_{1,1|0} = \hat{\mathbf{q}}_{1,0} = [-20,20]^{T}$ and $\bm{\Lambda}_{1,1} = \mathrm{diag}(0.01, 0.01)$, respectively.
Other system parameters are specified as $p_{1,1} = 0.1$ W, $b_{1,1} = 40$ MHz, and $H_{1} = 80$ m. 
First, it can be observed that the approximated OPs are generally larger than the Monte Carlo results in all cases, indicating that (\ref{fm:aOP-SG}) serve as an empirical upper bound of the ground-truth OP. 
Second, such empirical upper bound becomes tighter as the total number of IWSs $I_{\rm{g}}$ increases. 
Particularly, in the cases with $\epsilon_{\rm{s}} = 10^{-10}$ and $I_{\rm{a}},I_{\rm{r}}\leq 40$, the OP approximations are relatively inaccurate and fluctuate when their values are below $-5$ dB due to the coarse discretization of IWSs. 
In contrast, the OP approximation is satisfactorily close to the Monte Carlo results with a limited gap when $I_{\rm{a}}$ and $I_{\rm{r}}$ both increase to $100$.
As $I_{\rm{a}}$ and $I_{\rm{r}}$ further increase to $400$, the approximated OPs match the Monte Carlo results very well, which validates the accuracy of the IWS-based OP approximation when the radial and angular intervals $\Delta r^{\rm{s}}$, $\Delta \theta^{\rm{s}}$ are sufficiently small. 
Third, the improvement in approximation accuracy arising from increasing $I_{\rm{r}}$ from $20$ to $40$ with $I_{\rm{a}} = 20$ is relatively larger than that brought by further increasing $I_{\rm{a}}$ from $20$ to $40$ with $I_{\rm{r}} = 40$.
Finally, reducing $\epsilon_{\rm{s}}$ from $10^{-10}$ to $10^{-20}$ does not further improve the OP approximation accuracy in the case with $I_{\rm{r}} = I_{\rm{a}} = 40$. 
This is because for a fixed number of radial and angular intervals, decreasing $\epsilon_{\rm{s}}$ enlarges the radial and angular intervals, i.e., $\Delta r^{\rm{s}}$ and $\Delta \theta^{\rm{s}}$, according to (\ref{fm:dr}), and the coarser discretization degrades the approximation accuracy.  
Thus, merely decreasing $\epsilon_{\rm{s}}$ cannot improve the OP approximation accuracy, while a finer discretization of IWSs along both the radial and angular dimensions is more effective. 

\begin{figure}[!t]
    \centering
    \includegraphics[width=0.38\textwidth]{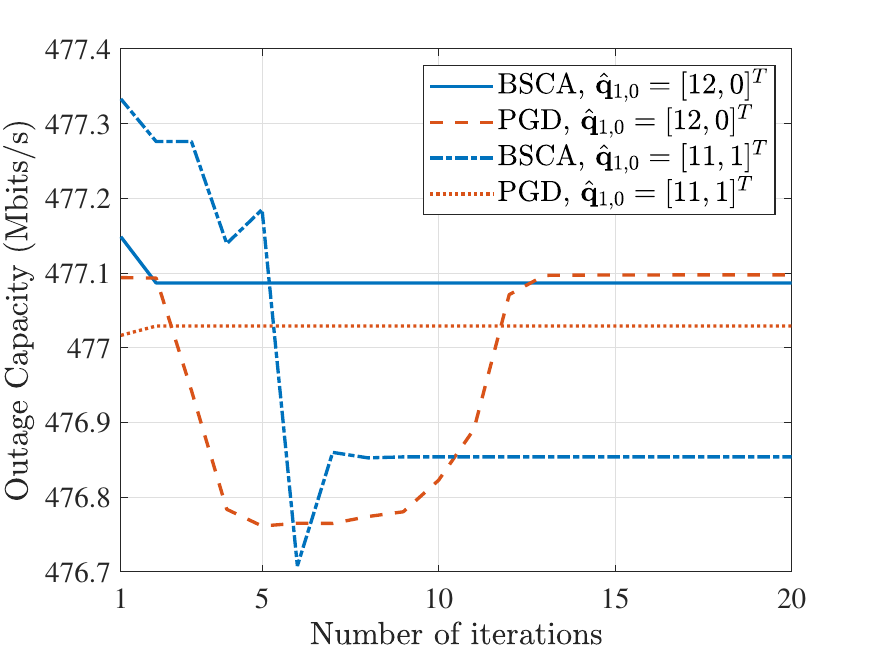}
    \caption{Convergence behaviour of the subalgorithms for predicted UAV position optimization.}
    \label{fig:A-CB}
    \vspace{-3mm}
\end{figure}

Before illustrating the optimization results of the two proposed schemes, the convergence behaviours of the proposed algorithms for both schemes are shown in Fig. \ref{fig:A-CB}. 
Thanks to the consistent form of the convex resource optimization subproblem for both schemes, it suffices to compare only the subalgorithms for predicted UAV position optimization. 
In particular, two cases with different estimated UAV positions $\hat{\mathbf{q}}_{1,0}$ are considered at a typical time slot $n=0$. 
In the case with $\hat{\mathbf{q}}_{1,0} = [12,0]^{T}$, the proposed PGD-based subalgorithm for the IWS-based scheme converges after almost 13 iterations, while the proposed BSCA-based subalgorithm for the CE-based scheme converges much faster. 
However, the convergence behaviours of the proposed subalgorithms in the case with $\hat{\mathbf{q}}_{1,0} = [11,1]^{T}$ are basically opposite to those in the case with $\hat{\mathbf{q}}_{1,0} = [12,0]^{T}$. 
Overall, the convergence of both subalgorithms is verified in Fig. \ref{fig:A-CB}, but their convergence speeds depend on the specific UAV position.

\begin{figure}[!t]
	\centering
    \subfigure[$\hat{x}_{1,0} = 0, \hat{y}_{1,0} = 20$.]{
        \vspace{-1mm}
        \label{fig:A-COR-1}
		\includegraphics[width=0.38\textwidth]{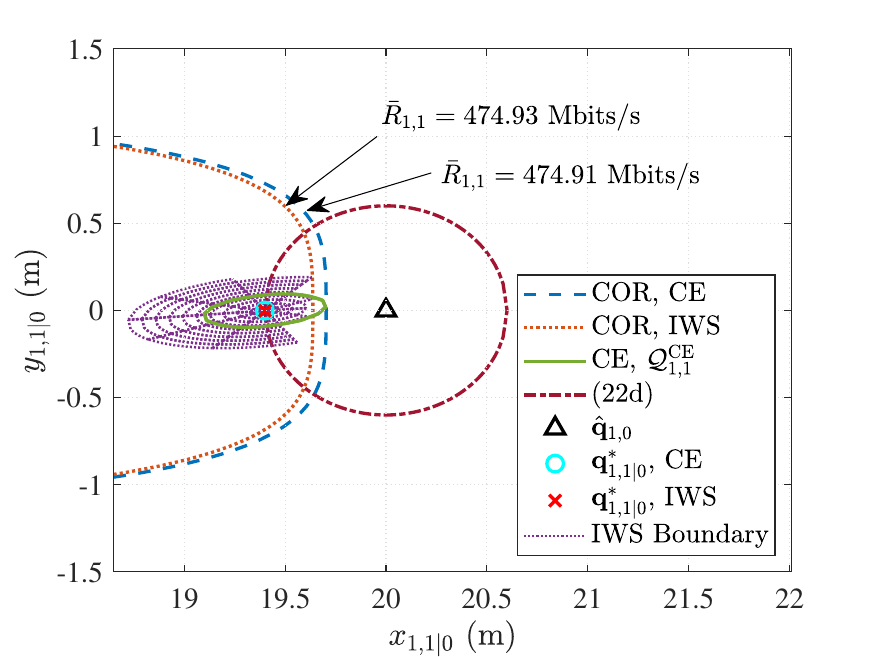}
	} 
    \subfigure[$\hat{x}_{1,0} = 20, \hat{y}_{1,0} = 0$.]{
		\vspace{-1mm}
        \label{fig:A-COR-2}
		\includegraphics[width=0.38\textwidth]{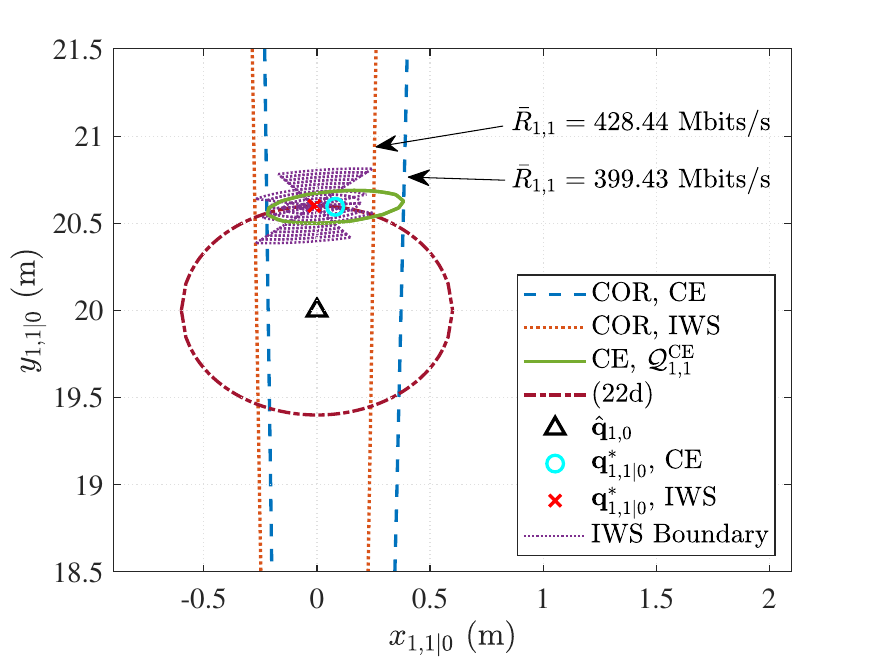}
    } 
    \label{fig:A-COR}
    \caption {Comparisons of optimized predicted UAV positions. }
    \vspace{-5mm}
\end{figure}

\subsection{Single-UAV Case} 

In this subsection, single-UAV cases with $k=1$ are studied to emphasize the performance gain achieved by UAV trajectory optimization of the proposed schemes. 
In these cases, the effectiveness of predicted UAV position optimization in the proposed schemes is demonstrated by comparisons with each other and with benchmark schemes.

We first compare the optimized predicted UAV positions and maximized outage capacities obtained by the two proposed schemes at a typical time slot $n=1$. 
Fig. \ref{fig:A-COR-1} and Fig. \ref{fig:A-COR-2} illustrate the optimized predicted UAV positions, CORs, and constructed uncertainty sets (i.e., the CE and IWSs) in two cases with different estimated initial UAV positions set as $\hat{\mathbf{q}}_{1,0} = [20, 0]^{T}$ and $\hat{\mathbf{q}}_{1,0} = [0, 20]^{T}$, respectively.
In both cases, the elements of the position prediction MSE matrix $\bm{\Lambda}_{1,1}$ are set as $[\bm{\Lambda}_{1,1}]_{11} = 0.01$, $[\bm{\Lambda}_{1,1}]_{12}=[\bm{\Lambda}_{1,1}]_{21} = 0.001$, and $[\bm{\Lambda}_{1,1}]_{22} = 0.001$.
Also, the boundaries of IWSs contained in the COR of the IWS-based scheme are shown in Fig. \ref{fig:A-COR-1} and Fig. \ref{fig:A-COR-2}.\footnote{The boundaries of every 10 IWSs are combined along each dimension since it is difficult to completely show the boundaries of $100\times 100$ IWSs.}
As shown in Fig. \ref{fig:A-COR-1}, the two proposed schemes result in very close optimized predicted UAV positions $\mathbf{q}_{1,1|0}^{*}$ and maximized outage capacities.
Nevertheless, in Fig. \ref{fig:A-COR-2}, the optimized predicted UAV positions of the two proposed schemes are different, and the maximized outage capacity of the IWS-based scheme is noticeably larger than that of the CE-based scheme. 
This observed difference is because containing the CE $\mathcal{Q}_{1,1}^{\rm{CE}}$ in the COR has different influences on the outage capacity in the two cases. 
Particularly, as shown in Fig. \ref{fig:A-COR-1}, the CORs of the two proposed schemes are only slightly different at their ends and are both sufficiently wide along the $y$-axis to contain the CE in the case with $\hat{\mathbf{q}}_{1,0} = [20,0]^{T}$. 
This slight COR difference leads to approximately the same outage capacity due to the COR definition (\ref{fm:COR}). 
However, in the case with $\hat{\mathbf{q}}_{1,0} = [0,20]^{T}$, the COR of the CE-based scheme is wider than that of the IWS-based scheme to contain the CE $\mathcal{Q}_{1,1}^{\rm{CE}}$, thereby leading to considerably different CORs and outage capacities as shown in Fig. \ref{fig:A-COR-2}.
Comparatively, the narrower COR of the IWS-based scheme is allowed by effectively utilizing the probability distribution of ground-truth UAV positions rather than relying solely on $\mathcal{Q}_{1,1}^{\rm CE}$, thereby rendering the IWS-based scheme superior to the CE-based scheme. 
Since both the CE and the probability distribution of ground-truth UAV positions are parameterized by the position prediction MSE matrix, whether the two proposed schemes result in substantially different maximized outage capacities depends on the UAV position and the position prediction MSE. 
In general, the proposed IWS-based scheme is more effective for robust trajectory optimization. 

Next, in a single-UAV case with $p_{\rm{tot}} = 0.1$ W and $b_{\rm{tot}} = 40$ MHz, the UAV trajectories optimized by the proposed schemes are compared with those obtained by the following benchmark schemes:
\begin{itemize}
    \item Rate maximization (R-Max): Assuming full transmit array gain, the predicted UAV position is optimized to maximize a predicted achievable rate denoted by 
    \vspace{-1.5mm}
    \begin{equation}
        \breve{R}_{1,n} = b_{\rm{tot}}\log_{2}(1+\frac{p_{\rm{tot}}\beta_{\rm{c}}N_{\rm{t}}}{N_{\rm{c}}b_{\rm{tot}}(\|\mathbf{q}_{1,n|n-1}\|^{2}+H_{1}^{2})}), 
    \vspace{-1.5mm}
    \end{equation}
    i.e., $\mathbf{q}_{1,n|n-1}$ is optimized to minimize the path loss subject to (\ref{opt-cstrt-d}).
    \item PCRB minimization (PCRB-Min): The predicted UAV position is optimized to minimize the sum of the PCRBs for the estimated UAV motion state \cite{YFJ2024CL}.
    The optimization problem is formulated as 
    \vspace{-1.5mm}
    \begin{equation}
        \min_{ \mathbf{q}_{1,n|n-1}  } \mathrm{Tr}(\mathbf{M}_{1,n}|_{\mathbf{x}_{1,n} = \mathbf{x}_{1,n|n-1}}) \
        \text{s.t.} \ 
        \text{(\ref{opt-cstrt-d})}. \notag  
        \vspace{-1.5mm}
    \end{equation}
\end{itemize}

\begin{figure}[!t]
    \centering
    \includegraphics[width=0.38\textwidth]{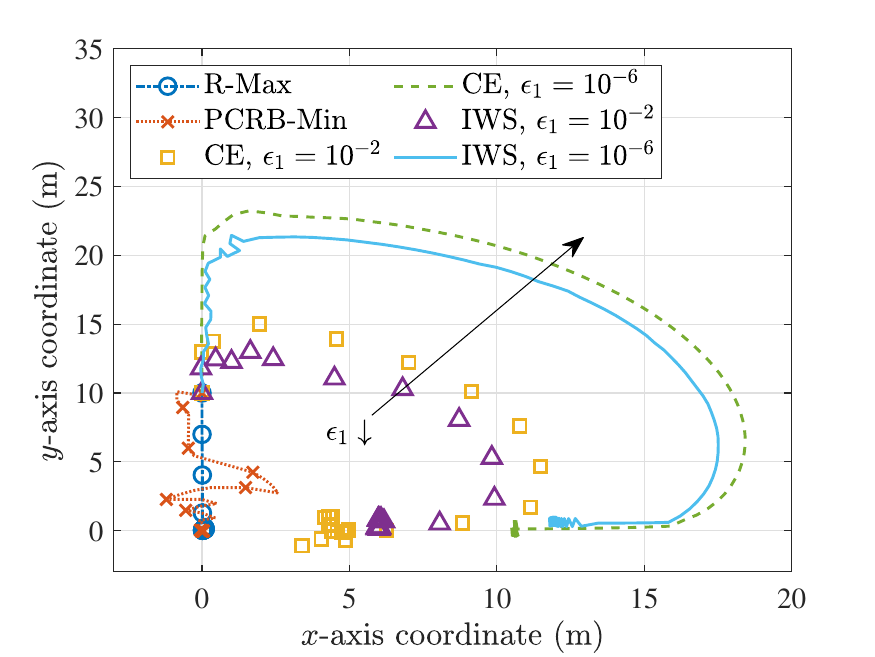}
    \caption{UAV trajectory comparisons in a single-UAV case.}
    \label{fig:B-Traj}
    \vspace{-3mm}
\end{figure}

Fig. \ref{fig:B-Traj} illustrates the optimized UAV trajectories versus time of the benchmark and proposed robust optimization schemes with $\epsilon_{1} = 10^{-2}$ and $\epsilon_{1} = 10^{-6}$ during $T=2.5$ s. 
The initial ground-truth and estimated UAV motion states in all cases are set as $\mathbf{x}_{0} = [0, 0, 10, 0]^{T}$ and $\hat{\mathbf{x}}_{0} \sim \bm{\mathcal{N}}(\mathbf{x}_{0},\mathbf{M}_{0})$ with $\mathbf{M}_{0} = \mathrm{diag}(10^{-3}, 10^{-4}, 10^{-3}, 10^{-4})$, respectively. 
Besides, the UAV trajectories of the proposed schemes in the case with $\epsilon_{1} = 10^{-2}$ are picked every 0.1 s.
The other system parameters are specified as $a^{\rm{az}}=0.01$ and $H_{1} = 80$ m. 
As shown in Fig. \ref{fig:B-Traj}, the R-Max scheme results in the UAV flying straight towards the head of the BS (i.e., $[0,0]^{T}$), while a zigzag UAV trajectory is obtained by the PCRB-Min scheme because its location need to be dynamically adjusted to reduce the measurement variances $(\sigma_{k,n}^{\text{az}})^{2}$ and $(\sigma_{k,n}^{\text{dst}})^{2}$ \cite{YFJ2024CL}.
In comparison, in both cases with $\epsilon_{1} = 10^{-2}$ and $\epsilon_{1} = 10^{-6}$, both of the proposed robust optimization schemes prevent the UAV from flying towards the head of the BS owing to the difficulty of vertical beam coverage for this area.
Instead, the UAV bypasses the neighbouring region around the head of the BS and flies towards the lateral side of the BS. 
The similar movement trends achieved by the CE-based and IWS-based schemes demonstrate their effectiveness in robust trajectory optimization.
Furthermore, compared with the case with $\epsilon_{1} = 10^{-2}$, the UAV trajectories of both proposed schemes are generally farther away from the BS in the case with $\epsilon_{1} = 10^{-6}$. 
Particularly, the UAV even increases its distance from the BS at the beginning of its movement in this case.
Therefore, maintaining a larger distance from the head of the BS is generally more advantageous for enhancing UAV communication reliability. 

\begin{figure}[!t]
    \centering
    \includegraphics[width=0.38\textwidth]{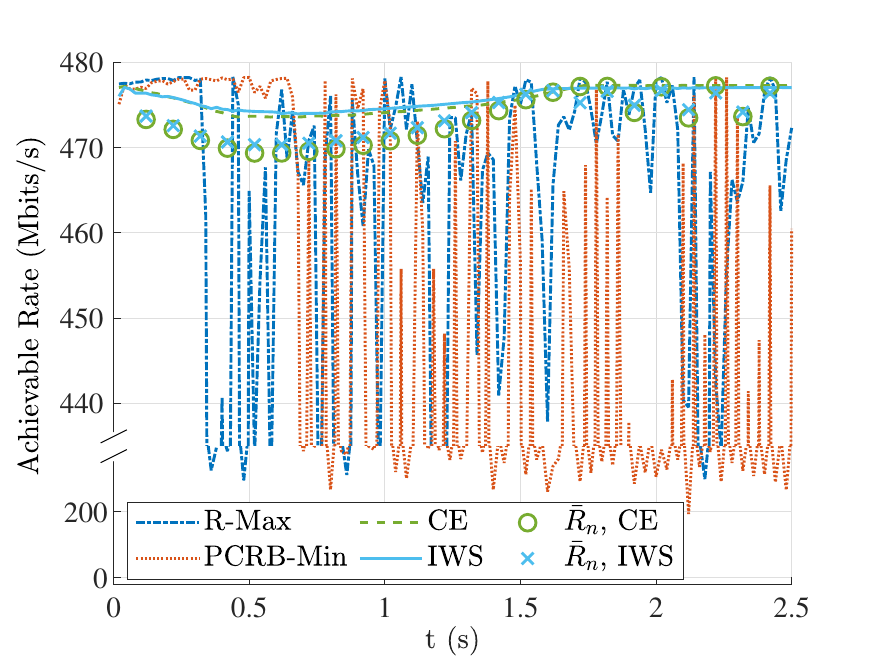}
    \caption{Rate comparisons between the proposed and benchmark schemes.}
    \label{fig:B-Rate}
\end{figure}

\begin{figure}[!t]
    \centering
    \includegraphics[width=0.38\textwidth]{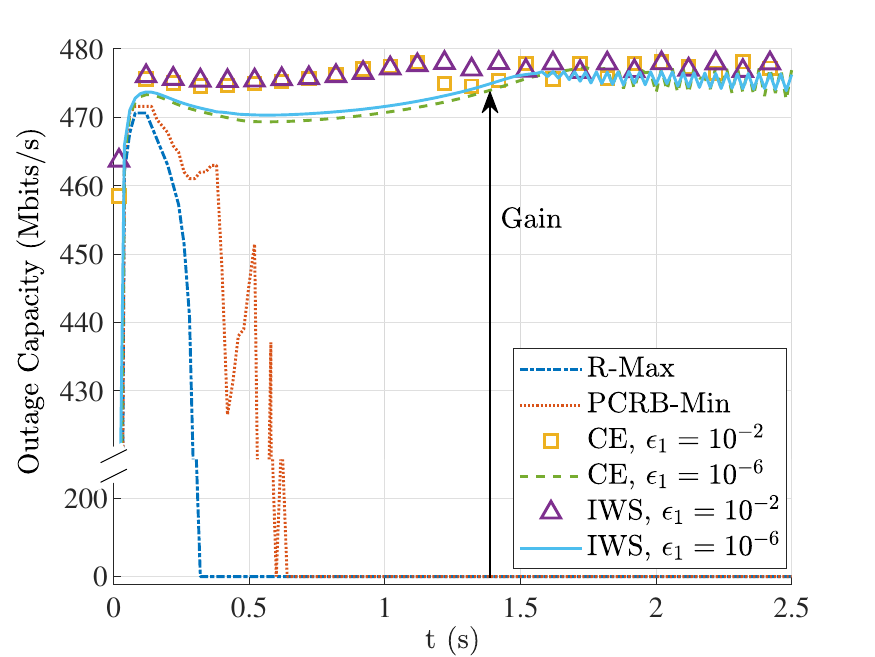}
    \caption{Outage capacity comparisons between the proposed and benchmark schemes.}
    \label{fig:B-bR}
    \vspace{-3mm}
\end{figure}

Fig. \ref{fig:B-Rate} shows the instantaneous achievable rates corresponding to the UAV trajectories in Fig. \ref{fig:B-Traj}, along with the outage capacities with $\epsilon_{1} = 10^{-6}$ obtained by our proposed schemes. 
It can be observed that the achievable rates resulting from both benchmark schemes fluctuate drastically between 200 and 478 Mbits/s, indicating their highly unstable communication performance owing to their improper UAV trajectory designs ignoring communication reliability. 
Comparatively, both proposed schemes lead to consistently high achievable rates, exhibiting their necessity for reliable communication and their significantly superior communication reliability to the benchmark schemes. 
Besides, the achievable rates are always larger than the maximized outage capacities, which verifies that the outage capacity provides a conservative and effective characterization of reliable communication performance. 

To further quantify the reliable communication performance enhancement arising from the proposed schemes, the outage capacities of the benchmark and proposed schemes are compared as shown in Fig. \ref{fig:B-bR}. 
Particularly, the UAV outage capacities of the benchmark schemes at the $n$-th time slot are calculated using the inverse cumulative distribution function of the achievable rate based on the verified OP approximation (\ref{fm:aOP-SG}) given the predicted UAV position. 
It can be observed that the outage capacities of the benchmark schemes decrease drastically to zero after the UAV approaches the head of the BS, indicating that the maximum tolerable OP constraint cannot be satisfied by any positive $\bar{R}_{n}$. 
In comparison, both of the proposed schemes maintain high outage capacities during the whole UAV flight in the case with $\epsilon_{1}=10^{-2}$, achieving a significant reliable data rate improvement over the benchmark schemes. 
Additionally, the outage capacities of proposed schemes in the case with $\epsilon_{1}=10^{-6}$ are generally lower than those in the case with $\epsilon_{1}=10^{-2}$, indicating a trade-off between enhancing the communication rate and reducing the OP. 

\subsection{Multi-UAV Case} 

\begin{figure}[!t]
    \centering
    \includegraphics[width=0.38\textwidth]{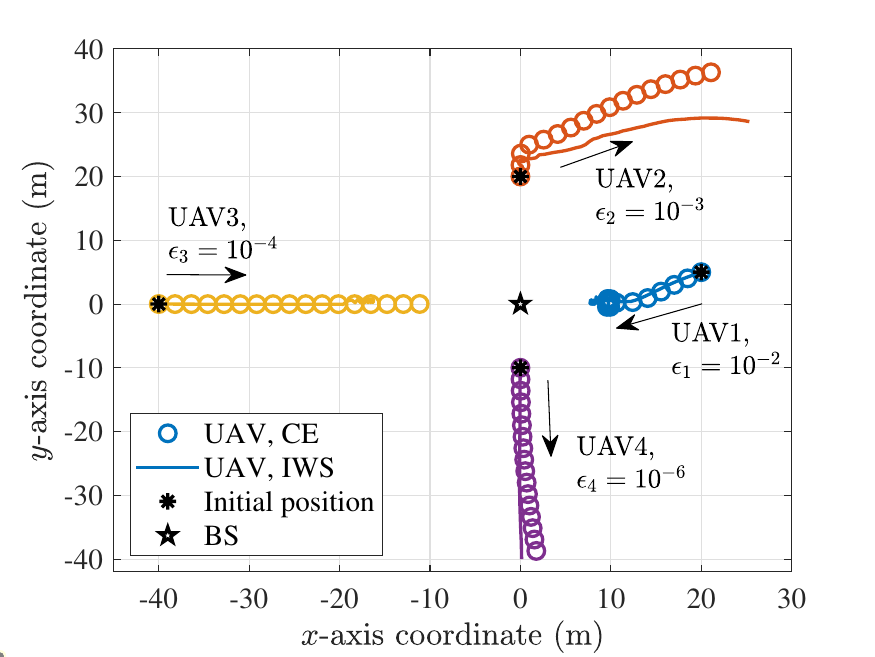}
    \caption{UAV trajectory comparisons in the considered multi-UAV case.}
    \label{fig:C-Traj}
    \vspace{-3mm}
\end{figure}

In this subsection, a multi-UAV case is studied mainly to illustrate the outage capacity improvement resulting from the resource optimization of the proposed schemes over the benchmark schemes. 
Specifically, the total number of UAVs is given by $K=4$ and the initial estimation MSE matrix of all UAVs is identically set as $\mathbf{M}_{0}=\mathrm{diag}(10^{-3},10^{-4},10^{-3},10^{-4})$.
The total power and bandwidth budgets are set as $p_{\rm{tot}} = 1$ W and $b_{\rm{tot}} = 100$ MHz, respectively. 
The other system parameters are specified as $a^{\rm{az}}=1$, $p_{\rm{min}}/p_{\rm{tot}} = b_{\rm{min}}/b_{\rm{tot}} = 0.1$, $p_{\rm{max}}/p_{\rm{tot}} = b_{\rm{max}}/b_{\rm{tot}} = 0.9$, $T=1$ s, and $H_{k} = 80$ m, $k\in\{1,2,3,4\}$. 
In this case, our proposed schemes are compared with two benchmark schemes for resource optimization described as follows:
\begin{itemize}
    \item R-Max: Define a predicted achievable rate for the $k$-th UAV at the $n$-th time slot by
    \vspace{-1.5mm}
    \begin{equation}
        \!\!\!\!\breve{R}_{k,n} \triangleq b_{k,n}\log_{2}(1+\frac{ p_{k,n}\beta_{\rm{c}}N_{\rm{t}} }{b_{k,n} N_{\rm{c}} (\|\mathbf{q}_{k,n|n-1}\|^{2} + H_{k}^{2})}). 
        \vspace{-1.5mm}
    \end{equation} 
    Then, the BS transmit power and bandwidth allocation are optimized by solving a predicted achievable rate maximization problem formulated as 
    \vspace{-1.5mm}
    \begin{equation}
        \max_{ \mathbf{p}_{n}, \mathbf{b}_{n} } \ \min_{k\in\{1,2,3,4\}}  \breve{R}_{k,n} \ 
        \text{s.t.} \ 
        \text{(\ref{opt-cstrt-a})-(\ref{opt-cstrt-c})}. \notag 
        \vspace{-1.5mm}
    \end{equation}
    \item PCRB-Min: The power and bandwidth are allocated by minimizing the maximum sum-PCRB for all UAVs, where the optimization problem is formulated as \cite{fwd}
    \vspace{-1.5mm}
    \begin{equation}
        \!\!\!\!\!\!\min_{ \mathbf{p}_{n}, \mathbf{b}_{n} } \ \max_{k\in\{1,2,3,4\}} \mathrm{Tr}(\mathbf{M}_{k,n}|_{\mathbf{x}_{k,n} = \mathbf{x}_{k,n|n-1}})  \ 
        \text{s.t.} \ 
        \text{(\ref{opt-cstrt-a})-(\ref{opt-cstrt-c})}. \notag 
        \vspace{-1.5mm}
    \end{equation}
\end{itemize}
To focus on the resource optimization comparison, the predicted UAV positions of benchmark schemes are the same as those of the IWS-based scheme. 
Also, the outage capacities of benchmark schemes are calculated as in the single-UAV case.

In the studied case, the UAVs are with different initial positions and maximum tolerable OP thresholds, which are detailed in Fig. \ref{fig:C-Traj} together with the optimized UAV trajectories of the proposed schemes. 
It can be noted from Fig. \ref{fig:C-Traj} that whether the UAV moves toward or away from the BS depends on both initial position and its maximum tolerable OP. 
Besides, the two proposed schemes can yield different real-time UAV trajectories due to different resource optimization results despite their similar movement trends, as shown in the case with $\epsilon_{2} = 10^{-3}$.  

\begin{figure}[!t]
	\centering
    \subfigure[Power allocation.]{
        \vspace{-1mm}
        \label{fig:C-P}
		\includegraphics[width=0.38\textwidth]{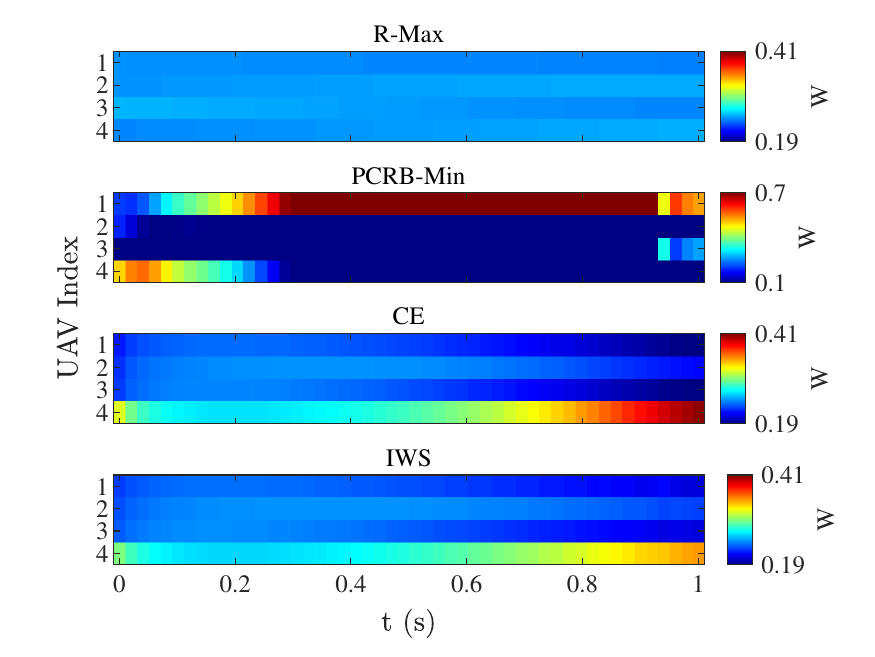}
	} 
    \subfigure[Bandwidth allocation.]{
		\vspace{-1mm}
        \label{fig:C-BW}
		\includegraphics[width=0.38\textwidth]{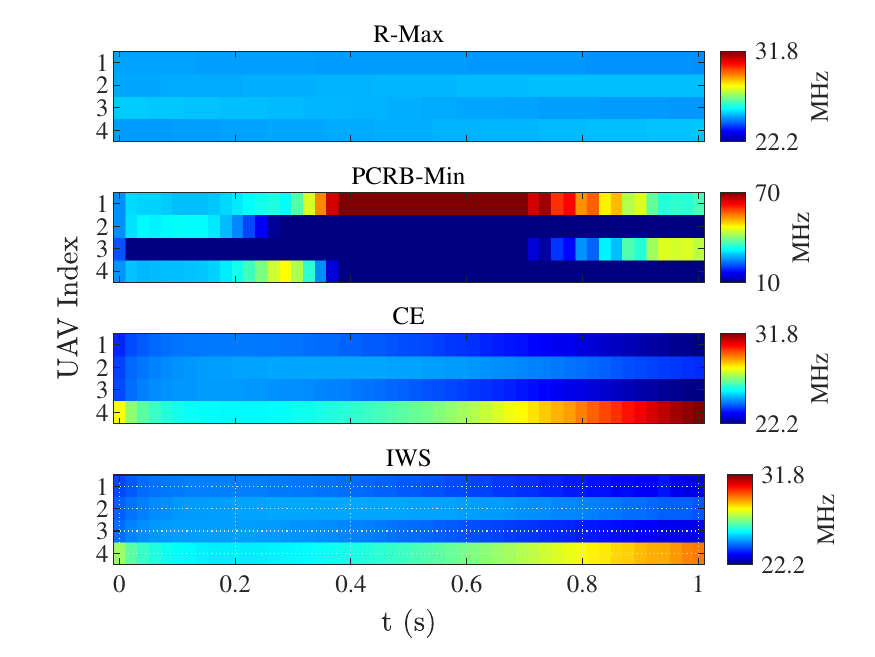}
    } 
    \label{fig:C-RA}
    \caption {Optimized resource allocation of all schemes. }
    \vspace{-5mm}
\end{figure}

Fig. \ref{fig:C-P} and Fig. \ref{fig:C-BW} illustrate the optimized power and bandwidth allocations of all schemes, respectively. 
By comparing Fig. \ref{fig:C-P} and Fig. \ref{fig:C-BW}, it can be seen that the time-varying pattern of power allocation is similar to that of bandwidth allocation for each scheme. 
Specifically, more power and bandwidth are allocated by the R-Max scheme to the UAVs with larger distances from the BS than the other UAVs. 
Since the UAVs have only limited differences in their distances to the BS, the R-Max scheme results in an approximately equal resource allocation among UAVs. 
However, different from that, the PCRB-Min scheme can lead to highly uneven power and bandwidth allocation among UAVs. 
The temporal variations of resource allocations resulting from the PCRB-Min scheme are because the measurement covariance matrix $\mathbf{Q}_{k,n}^{\rm{m}}$ incorporated in the sum-PCRB depends on both resource allocation and UAV positions. 
The optimized resource allocations of benchmark schemes show little dependence on the maximum tolerable OP for each UAV, i.e. $\epsilon_{k}$. 
In contrast, the two proposed schemes allocate the fewest and most resources to the UAVs with the largest and smallest $\epsilon_{k}$, respectively, which demonstrates that both proposed schemes are aware of outage constraints for UAVs. 
Besides, the time-varying patterns of resources allocated to UAV 1--3 under the two proposed schemes are similar to those under the R-Max scheme, thereby achieving a trade-off between the communication efficiency and reliability of the overall system. 
Furthermore, despite the similar time-varying patterns of optimized resource allocations, the IWS-based scheme allocates less BS transmit power and bandwidth to UAV 4 than the CE-based scheme, yielding more balanced resource allocations among the UAVs to satisfy the outage constraints. 

\begin{figure}[!t]
    \centering
    \includegraphics[width=0.38\textwidth]{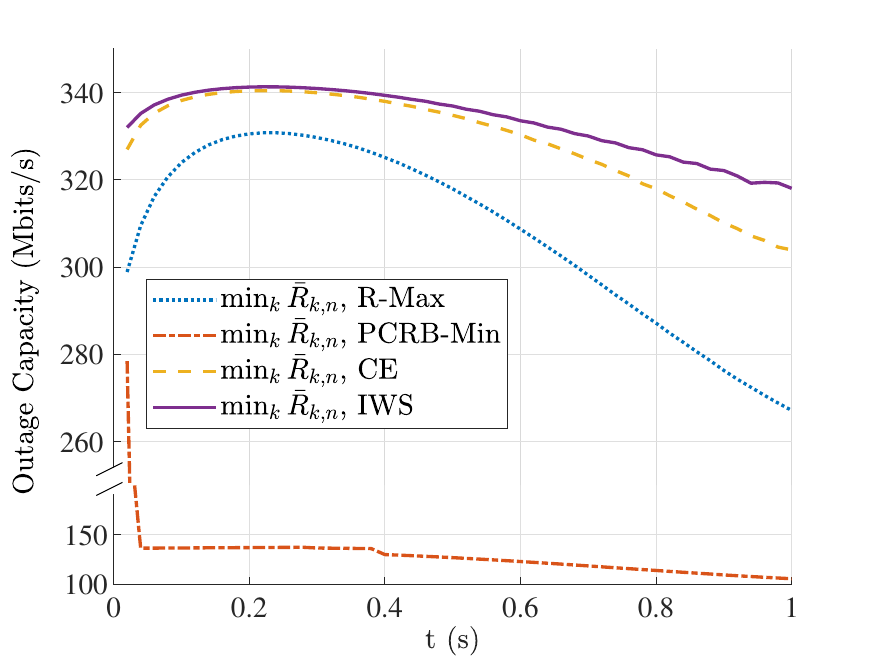}
    \caption{Outage capacities of all schemes for resource allocation.}
    \label{fig:C-OC}
    \vspace{-3mm}
\end{figure}

The minimum outage capacities among UAVs of the benchmark and proposed schemes are compared and illustrated in Fig. \ref{fig:C-OC}.
It can be observed that the two proposed schemes lead to a significant outage capacity gain over the benchmark schemes. 
The extremely low minimum outage capacity among UAVs of the PCRB-Min scheme is mainly caused by its uneven resource allocations as shown in Fig. \ref{fig:C-RA}, while the improvement of the proposed schemes over the R-Max scheme is because of the more allocated power and bandwidth to the UAV with the strictest maximum tolerable OP constraint. 
Compared to the CE-based scheme, the IWS-based scheme achieves higher and more stable outage capacities because of its more balanced resource allocations as shown in Fig. \ref{fig:C-RA}. 
Therefore, the IWS-based scheme yields more effective resource optimization than the CE-based scheme for outage capacity maximization. 

\section{Conclusions}

In this paper, we proposed to maximize the minimum outage capacity among multiple cellular-connected UAVs by optimizing the real-time predicted UAV positions, BS transmit power allocation, and bandwidth allocation for reliable sensing-aided communication. 
To deal with the non-convex and implicit maximum tolerable OP constraints, two robust optimization schemes were proposed by forcing the COR to contain an uncertainty set constructed from a continuous CE and discretized IWSs, respectively.
The CE-based scheme optimized the predicted position of each UAV based on the BSCA technique, while the IWS-based scheme utilized a proposed IWS selection to decouple the optimization variables and optimized the predicted UAV positions based on the PGD approach. 
Both schemes subsequently solve a common convex resource allocation subproblem and are further unified into a sequential optimization framework, revealing that robust UAV trajectory and resource optimization can be effectively decomposed through proper uncertainty set construction. 
Simulation results verified the accuracy of the proposed IWS-based OP approximation and demonstrated the effectiveness of the proposed robust optimization schemes, which achieved significant outage capacity improvement over the benchmark schemes dedicated for sensing or communication. 
Furthermore, the IWS-based scheme was shown to outperform the CE-based scheme in robust trajectory and resource optimization. 
The extension of our proposed robust optimization schemes and framework to multi-BS cooperative scenarios is an interesting direction for future work.

\bibliographystyle{IEEEtran}
\bibliography{IEEEabrv,ref}

\end{document}